\documentclass{elsart}
\usepackage{graphicx}
\usepackage{lscape}
\def\contrac#1#2#3<#4>{%
\setbox1=\hbox{$#1$}%
\setbox2=\hbox{$#2$}%
\setbox3=\hbox{$#3$}%
\dimen0=.5\wd1%
\advance\dimen0 by \wd2%
\advance\dimen0 by .5\wd3%
\setbox0=\hbox{\kern.5\wd1%
\vtop{\hbox to \dimen0{\vrule depth#4\hfil\vrule}\hrule}}%
\dimen0=#4 \advance\dimen0 by \lineskip \dp0=\dimen0%
\wd0=0pt%
\setbox1=\vtop{\lineskiplimit=10cm\lineskip=1mm \box1\box0}%
\box1\box2\box3}
\def\nuc#1#2{\relax\ifmmode{}^{#1}{\protect\text{#2}}\else${}^{#1}$#2\fi}
\def\nucZ#1#2#3{\relax\ifmmode{}_{#1}^{#2}{\protect\text{#3}}\else$_{#1}^{#2}$#3\fi}
\begin{document}
\begin{frontmatter}
\title{ Coulomb exchange and pairing contributions in nuclear
Hartree--Fock--Bogoliubov calculations with the Gogny force}
\author{ M. Anguiano, J.L. Egido and L.M. Robledo }
\address{ Departamento de F\'{\i}sica
Te\'orica, Universidad Aut\'onoma de Madrid,  \\
      E-28049 Madrid, Spain }

\date{\today}
\maketitle
\begin{abstract}
We present exact Hartree--Fock--Bogoliubov  calculations with the finite range 
density dependent Gogny force using a triaxial basis. For the first time, 
all  contributions  to the Pairing  and Fock Fields arising from the Gogny and 
Coulomb interactions as well as the two-body correction of the kinetic energy 
have been calculated in this basis. We analyze the relevance of these terms 
in different regions of the periodic table at zero and high angular momentum. 
The validity of commonly used approximations that neglect different terms in 
the variational equations is also checked.  We find a decrease of the proton
pairing energies mainly due to a Coulomb antipairing effect.

\end{abstract}
\end{frontmatter}
\section{Introduction}

The mean field  approximation is the backbone of many-body 
calculations in Nuclear Physics, either as a zero order approach to the problem 
or as a basis for more elaborated theories like the Random-Phase approximation 
or the Generator Coordinate Method.  The  modern and fast computer facilities
have made possible the use of effective density dependent interactions, 
as the Skyrme \cite{Vau72} or the Gogny force \cite{Dec80}, in standard 
mean field calculations like the Hartree-Fock (HF) or the 
Hartree-Fock-Bogoliubov (HFB) approaches.

 In spite of the relative simplicity of the mean field approach, some additional
approximations  have been commonly used when effective forces were used  
in the mean field calculations in order to make the problem more tractable. 
The complications in 
the calculations usually arise from the exchange terms, either  from the 
Coulomb force or from some components of the nuclear force itself.
 Depending on the mean field approach 
(HF or HFB)  and on the force (zero or finite range) different approximations 
have been used in the past. In the HF theory one has only one exchange term 
(the Fock term) which contributes to the HF field.  Commonly, in this type of
calculations, in order to retain the required simplicity of the model, 
the Fock term  of the Coulomb force has been neglected, 
or treated in the Slater approximation. In ref.~\cite{Tit74} the validity of 
different approaches for the Coulomb Fock contribution was tested
in HF calculations finding the Slater approximation  a good one.
In the HFB approach, and for any two-body interaction, one has two exchange 
terms, the Fock term  which contributes to the HF field and the 
particle-particle ({\it pp}) term which contributes to the pairing
field. In the HFB approach one must distinguish between the Skyrme and the Gogny 
force. Since the Skyrme force is assumed to provide only the particle-hole
 ({\it ph})
part of the force, all terms of the {\it pp} type are neglected even
those arising from the Coulomb force.
 With the Gogny force the situation is different because with its finite range
  it has been designed to 
provide both, the {\it ph} and the {\it pp} part of the nuclear force.

  A test similar to the one performed for HF in ref.~\cite{Tit74} has never 
been done for HFB calculations. Exact HFB calculations with the Gogny force
have been performed for spherical nuclei \cite{Dec80}, but the effects of the 
approximation of neglecting the exchange terms  never have been analyzed. For 
the  more complex 
triaxial HFB calculations it has been assumed that the findings of
ref. \cite{Tit74} for the HF approach, concerning the exchange terms, will 
also apply to this case. This point, however,  never has been checked.  

HFB calculations with  triaxial symmetry using density--dependent
interactions like the Gogny force, used by different 
groups~\cite{Gir83,Egi93,Gir94}, are usually performed using the following 
approximations: 
\begin{itemize}
\item First, the contributions to the pairing field stemming from the center
of mass correction, Coulomb and spin--orbit terms are neglected. 
This is motivated by the fact that their 
contribution, in the limit of axial deformations, is assumed (but not proved)
to be negligible small~\cite{Gir83}. Thus, under these conditions the pairing 
field arises only from the Brink--Boeker term. 
 \item Second, since  the exchange (Fock) Coulomb 
term contribution to the HF field requires a large  CPU time, its contribution
to the energy is not taken into account or it is included  in the Slater 
approximation. 
\end{itemize}
 
 On the other hand  in the HFB theory,  pairing is an important degree of 
freedom and it is important to analyze  how  the properties related to 
this part of  the field change when some approximations are assumed. 
In this paper, for the first time, we calculate  in the triaxial basis
the contributions mentioned above . 
We shall investigate  the effect of including all the terms in 
the fields as compared with the calculations which are normally 
done for deformed nuclei.
Mainly the D1S parameterization of the Gogny force~\cite{Ber91} is  used in
the calculations, though some calculations with the D1~\cite{Dec80,Gog75} 
parameterization  will be done. We shall study nuclei
which have been analyzed by the approximate triaxial HFB, like some nuclei 
in the rare earth region~\cite{Egi93,Egi94},  in the superdeformed  
A$\approx 190$ region~\cite{Gir94} and in the  actinide region~\cite{Gir83}. 
In the same way, we shall also analyze these contributions in spherical nuclei 
in order to understand how  the results obtained  depend on the basis used.
A brief review of the method is given in  Sect.~\ref{sec:Met}. 
The effect of the different contributions in spherical nuclei is studied 
in Sect.~\ref{sec:SN50} for the N$=50$ region and in Sect.~\ref{sec:SN82} 
for the N$=82$ y N$=126$ regions. The rare earth 
region will be analyzed in the Sect.~\ref{sec:REN}. The effect of the new 
terms in the nucleus \nuc{240}{Pu} (actinide region) will be treated in  
Sect.~\ref{sec:AR}.  
The behavior of a nucleus in the superdeformed 
region $A \approx 190$ will be studied in  Sect.~\ref{sec:SHg}. Finally an
 study of energy surfaces with considerations of the exchange terms will be done 
in Sec.~\ref{sec:enersurf}. 
 A discussion
and the conclusions will be presented at the end of the paper.

\section{Theory}
\label{sec:Met}

The HFB theory~\cite{Rin80} unifies the self--consistent description of nuclear
orbitals, as given by the Hartree--Fock (HF) approach and the
Bardeen--Cooper--Schrieffer (BCS) pairing theory into a single variational
theory. As trial wave function  an independent quasiparticle state
$| \Phi \rangle$ is considered. This wave function is a linear combination of 
independent multi-particle states representing various possibilities of occupying
single--particle states. 
  The quasiparticle operators are defined by
\begin{equation}  \label{bogtrans}
\alpha _\mu =\sum_iU_{i\mu }^{*}c_i+V_{i\mu }^{*}c_i^{\dagger},
\end{equation}
with ${c_{k}^{\dagger},c_{k}}$ the particle creation and annihilation 
operators in the harmonic oscillator basis and $U$ and $V$ the Bogoliubov 
wave functions to be determined by the Ritz variational principle. The
corresponding equations have been solved by the Conjugate Gradient Method
\cite{ELM.95}.
The Bogoliubov transformation (\ref{bogtrans}) leads to wave functions 
$|\Phi \rangle$
that are not eigenstates of the particle number operator. Furthermore, 
to generate a wave function 
$|\Phi_I \rangle$ corresponding to a rotational state with angular 
momentum $I$ we shall use the cranking prescription. Therefore, in the
minimization process one has to add Lagrange multipliers to satisfy the
pertinent constraints. That means, the wave 
function  $|\Phi_I \rangle$ is determined by the condition
\begin{equation}
\delta \langle \Phi_I | \hat{H} -\omega \hat{J}_x -\lambda_N \hat{N} -\lambda_Z
\hat{Z} | \Phi_I \rangle =0
\label{chfb}
\end{equation}
with the Lagrange multipliers $\omega, \lambda_N$ and $\lambda_Z$ determined 
by the usual constraints $\langle \hat{J}_x \rangle= \sqrt{I(I+1)}$, 
$\langle \hat{N} \rangle =N$ and $\langle \hat{Z} \rangle =Z$, in an obvious
notation.

In the second quantization formalism,  the nuclear Hamiltonian is given by
\begin{equation}
\hat{H}= \sum_{k_1k_2} t_{k_1k_2}c_{k_1}^{\dagger}c_{k_2} \, + \,
\frac{1}{2}
\sum_{k_1k_2k_3k_4}{v}_{k_1k_2k_3k_4}c_{k_1}^{\dagger}
c_{k_2}^{\dagger}c_{k_4}c_{k_3}. 
\end{equation}
where $t$ is the kinetic energy and $v$ the two-body interaction,
$\bar{v}$ if antisymmetrized.
  The Wick theorem  allows the evaluation of the expectation value
of a two-body operator in a simple way:
\begin{equation} \label{wick}
\langle c_{k_1}^{\dagger}c_{k_2}^{\dagger}c_{k_4}c_{k_3} \rangle = 
\langle c_{k_1}^{\dagger}c_{k_3} \rangle \langle c_{k_2}^{\dagger}c_{k_4} \rangle
\, - \,
\langle c_{k_2}^{\dagger}c_{k_3} \rangle \langle c_{k_1}^{\dagger}c_{k_4} \rangle
\, + \,
\langle c_{k_1}^{\dagger}c_{k_2}^{\dagger} \rangle \langle c_{k_4}c_{k_3}
\rangle,
\end{equation}
where we have introduced the shorthand notation $ \langle \Phi | \hat{O} | \Phi \rangle
\equiv \langle  \hat{O}  \rangle $ for the expectation value of any operator
$ \hat{O} $. The first term on the r.h.s. of eq.~(\ref{wick}) is called the
Hartree term, the second one the Fock term and the third one the pairing term.
Using the expressions 
\begin{equation} 
 \rho_{k_1,k_2} = \langle c_{k_2}^{\dagger}c_{k_1} \rangle,
 \;\;\;\; \kappa_{k_1,k_2} = \langle c_{k_2}c_{k_1} \rangle, 
\end{equation} 
for the density matrix $\rho$ and the pairing tensor $\kappa$, and
\begin{equation} \label{fields}
\Gamma_{k_1k_3} = \sum_{k_2k_4} \bar{v}_{k_1k_2k_3k_4}\rho_{k_4k_2}, \quad 
\Delta_{k_1k_2} = \frac{1}{2} \sum_{k_3k_4} \bar{v}_{k_1k_2k_3k_4}\kappa_{k_3k_4},
\end{equation}
for the HF field $\Gamma$ and the pairing field $\Delta$, we can express  
the expectation value of the nuclear Hamiltonian by
\begin{equation}
E= \langle \hat{H} \rangle =
         {\displaystyle Tr}  ( t \rho) +
\frac{1}{2}  {\displaystyle Tr}  \left (\Gamma \rho  \right) -
\frac{1}{2}  {\displaystyle Tr}  \left (\Delta {\kappa}^* \right).
\label{eq:EHFB}
\end{equation}
 The second addend in this expression arises from the
contributions of the Hartree and the Fock terms of eq.~(\ref{wick}) while the 
third one stems from the pairing term of eq.~(\ref{wick}).
The two-body effective nuclear interaction $v$ used in our calculations,
see Appendix~A, is 
made up from the Gogny force, the Coulomb (C) interaction and
the two--body correction of the kinetic energy (TK).
The Gogny force itself  has the following terms:
Brink--Boeker (BB), spin--orbit (SO) and density--dependent (DD) contributions. 
 Then, the total HFB energy can be written as
\begin{equation}
E= {\displaystyle Tr} (t \rho) + V_{BB} + V_{SO} + V_{DD} + V_{TK} + V_{C}
\label{hfbe}
\end{equation}
where we have written separately the different energy contributions. 
Each contribution, $V_L$, is  calculated from the corresponding Hartree--Fock,
$\Gamma_L$, and pairing, $\Delta_L$, fields (L $\equiv$ BB, SO, DD, TK or C) and
is given by
\begin{equation}
V_L = \frac{1}{2} {\displaystyle Tr} \, \left ( \Gamma_L \rho - 
\Delta_L \kappa^* \right ) = V_L^H + V_L^F + V_L^P
\label{VLterms}
\end{equation}
where $\Gamma_L$ and $\Delta_L$ are given by eq.~(\ref{fields}) but considering,
instead of the full interaction $v$, 
only the part $L$ of the force. On the r.h.s. of this expression we have
further separated $V_L$ into the three contributions obtained from each of the
terms of eq.~(\ref{wick}), namely, the Hartree-, the Fock- and the 
pairing term. 

As mentioned above, in HFB calculations with the Gogny force and a triaxial 
basis several approximations have been done. In particular the contributions 
to the pairing field  from the spin--orbit, two--body correction of the kinetic 
energy and  Coulomb terms have been usually neglected in the self--consistent
procedure. Furthermore, the exchange contribution of the Coulomb potential 
(Fock term) to the Hartree--Fock field has not been  taken into account in most
 of the cases or treated in the Slater approximation.
  Thus, in the simplest HFB approximation, which we shall denote 
HFB$_s$ ($s$ for standard), the HFB energy used in the variational process is 
given by
\begin{equation}
E_s = {\displaystyle Tr} (t \rho) + V_{BB} + V_{DD} + \frac{1}{2} 
{\displaystyle Tr}  \left ( \,( \Gamma_{SO} + 
\Gamma_{TK} + \Gamma^{H}_{C} ) \rho \, \right )
\label{hfba}
\end{equation}

A first order perturbation theory correction to this approach, which we
shall denote  HFB$_{s+}$, is to calculate the neglected terms just at the end 
of the iterative procedure and to add  them to the total energy. We shall
test, in different mass regions, how good this approach is with respect to the 
exact calculation (\ref{eq:EHFB}), denoted HFB$_{e}$ ($e$ for exact), and with
respect to other approaches that we shall introduce later on.
Details on the calculation of the terms neglected in the HFB$_{s}$ approximation
but included in the exact one are given in Appendix~B.

The triaxial basis used in the calculations is spanned by  harmonic oscillator  
states with quantum numbers $\{ n_x,n_y,n_z \}$ which fulfill the condition, 
$a_xn_x+a_yn_y+a_zn_z \leq N_0$. The coefficients $a_i$ are related
to the axis of the nuclear matter distribution by 
$a_x={(qp)}^{1/3}$, $a_y=q^{1/3}p^{-2/3}$ and
$a_z=p^{1/3}q^{-2/3}$~\cite{Gir83}.

\section{Spherical nuclei}

In this section we concentrate mainly on the effect of the Coulomb contribution
 in HFB calculations for spherical nuclei.  We are interested to know the
behavior of the pairing and total energies when different approximations for
the Coulomb terms are used. At the end of sect.~\ref{sec:SN50}, a short 
discussion is devoted to the contributions to the total energy of 
the pairing field  of both, the spin--orbit term and the two--body correction
 of the kinetic energy term.
\subsection{N$=50$ region}
\label{sec:SN50}
  In this subsection we shall study the medium heavy nuclei from
the  region $N=50$. We shall use the basis determined by $q=1.0$, $p=1.0$ 
and $N_0=11.1$. In the calculations the D1S parameterization of the
Gogny force has been used. First we analyze in detail the Coulomb terms and
ignore for the moment the SO and TK pairing exchange terms.

The Coulomb interaction contributes to the energy
with three terms~: The Hartree (H / h) or direct term, the Fock (F / f) or
 exchange term and the pairing (P / p) or Bogoliubov term. Each term can be 
considered either selfconsistently, i.e., taken into account in the variational process,
or non-selfconsistently (perturbationally), i.e., ignored during the variation 
but its value being added to the total energy at the end of the variational 
process. In the first case we shall use capital letters (H, F or P) and in 
the second one small letters (h, f or p) to label the approximation used
for the respective terms.
For example the approximation $Hfp$ means that the
Hartree Coulomb term has been taken into account selfconsistently and 
the Fock and Pairing Coulomb terms only perturbationally.
 To understand which contributions are more important and the convenience 
of considering in the variational process the pertinent terms we shall 
investigate the following approximations:

 In the simplest calculation the three Coulomb terms mentioned 
 above, i.e., even the Coulomb Hartree term, are neglected. In this
approximation the energy, which  we shall call reference energy, $E_{ref}$, 
 given by
\begin{equation}
E_{ref}  =  {\displaystyle Tr} (t \rho) + V_{BB} + V_{DD} + \frac{1}{2} 
{\displaystyle Tr}  \left ( \,( \Gamma_{SO} + 
\Gamma_{TK} ) \rho \, \right ) 
\label{eref}
\end{equation}
is considered in the minimization process.
 The next  approximations will be

\begin{eqnarray}
 E_{hfp} & =  & E_{ref} + [ V_{C} ] \\
E_{Hfp}  & =   & E_{ref} + V_{C}^{H} + [ V_{C}^{F} + V_{C}^{P} ] \\
E_{HFp}  & =   & E_{ref} + V_{C}^{H} + V_{C}^{F} + [ V_{C}^{P} ] \\
E_{HFP}  &  =   & E_{ref} + V_{C} \label{EHFP}
\end{eqnarray}

We use  squared brackets to indicate the terms considered perturbationally,
the others, are used in the variational equations. Notice, therefore, that only
the latter ones determine the wave function.   
 To remember the meaning of each approximation one just needs to look at the
subindex of the energy and to keep in mind the convention introduced above 
about capital and small letters. 
 We shall also investigate the Slater approximation used by 
several groups. In this  case the energy looks like
\begin{equation}
E_{H\tilde{F}p}  =   E_{ref} + V_{C}^{H} + \tilde{V}_{C}^{F} + [ V_{C}^{P} ]
\end{equation}
In $E_{H\tilde{F}p}$ the tilde on $F$ means that the Fock term has been
calculated in the Slater approximation but it has been considered in the 
variational process. In these calculations the best one is, obviously, $HFP$
where all Coulomb terms are taken into account selfconsistently.

\begin{figure}[h]
\begin{center}
\parbox[c]{14cm}{\includegraphics[width=14.cm]{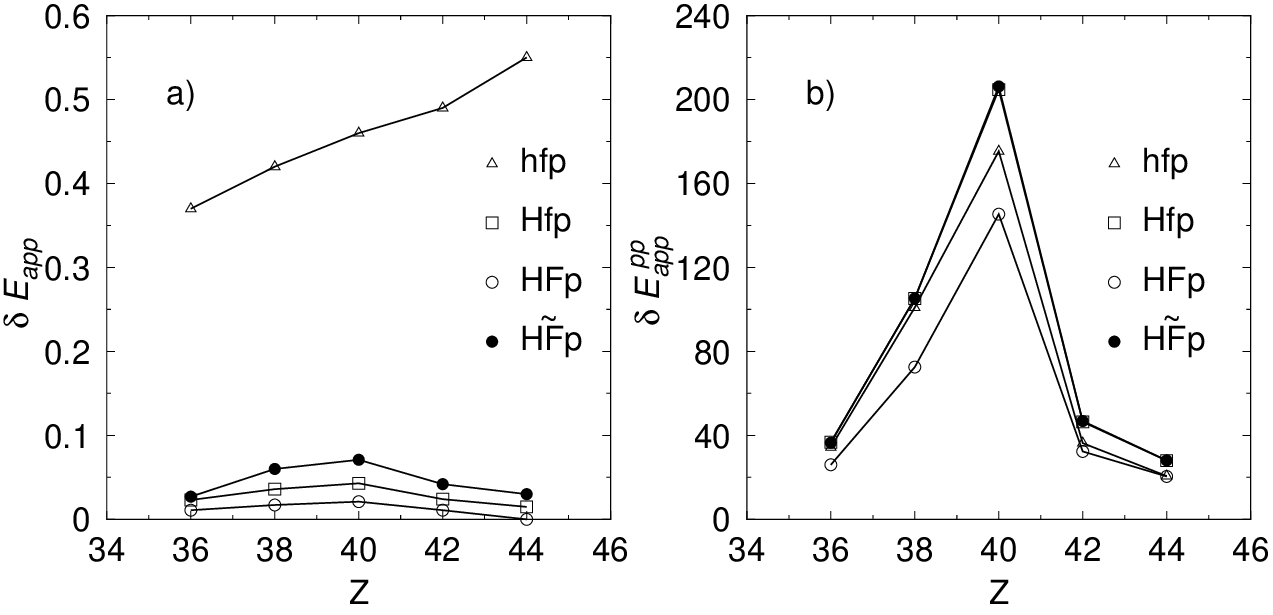}}
\end{center}
\caption{Percentage of error in a) the binding energy and b) the pairing energy 
of various approximations for the Coulomb part related to exact calculations 
as a function of the proton number Z.}
\label{fig:EPTSN}
\end{figure}

In the upper part of Table~\ref{tab:SN} we present the  binding energies
of five spherical nuclei with $N=50$ calculated in the different
approximations. The experimental binding energies are from ref.~\cite{Aud93}.
 As expected from a variational method, the more terms are treated
selfconsistently the deeper gets the energy minimum. The largest energy 
decrease 
is provided by the inclusion of the Hartree term in the variational equations.
The largest deviation of the $HFP$ energies from the experimental 
values is about 2 MeV, but already the selfconsistent treatment of
the Hartree term provides a good approximation to it.
We also find that the Slater approximation provides energies close to the
$Hfp$ approximation. These effects can be more clearly seen in 
Fig.~\ref{fig:EPTSN}a where we plot the quantity $\delta E_{app.}$ which 
represents the percentage of error of a given approximation with respect to the
 calculation $HFP$ and  is defined by  
\begin{equation}
   \delta E_{app.} = 100 \times \frac{E_{app.} - E_{HFP}}{E_{HFP}}
\label{percen}
\end{equation}  
where $E_{app.}$ represents any of the approximations that we have
considered. Here, we see that the $hfb$ approximation, where none of the 
Coulomb terms is treated  selfconsistently, deviates the most from the exact
 $HFP$ results and that the deviation increases with the proton number. The 
other approaches provide a good approximation to the complete $HFP$.

\begin{table}[h]
\begin{center}
\begin{tabular}{lccccc}
\hline 
  & \nucZ{36}{86}{Kr} & \nucZ{38}{88}{Sr} & \nucZ{40}{90}{Zr} & \nucZ{42}{92}{Mo} & 
\nucZ{44}{94}{Ru} \\
\hline
E$_{hfp}$          & -744.265 & -764.352  & -780.868  &  -793.678  &  -803.980 \\
E$_{Hfp}$          & -746.827 & -767.259  & -784.105  &  -797.425  &  -808.286 \\
E$_{H\tilde{F}p}$  & -746.801 & -767.093  & -783.887  &  -797.284  &  -808.175 \\
E$_{HFp}$          & -746.920 & -767.410  & -784.281  &  -797.529  &  -808.352 \\
E$_{HFP}$          & -747.002 & -767.539  & -784.446  &  -797.616  &  -808.411 \\
E$^{exp}_{tot}$    & -749.231 & -768.464  & -783.894  &  -796.509  &  -806.843 \\
\hline
E$^{pp}_{ref}$          & -10.332 &   -8.092 & -6.967 & -8.644  &  -9.214 \\
E$^{pp}_{hfp}$          & -9.352  &   -7.335 & -6.314 & -7.798  &  -8.278 \\
E$^{pp}_{Hfp}$          & -9.522  &   -7.490 & -6.988 & -8.377  &  -8.785 \\
E$^{pp}_{H\tilde{F}p}$  & -9.509  &   -7.494 & -7.020 & -8.396  &  -8.791 \\
E$^{pp}_{HFp}$          & -8.772  &   -6.297 & -5.627 & -7.567  &  -8.270 \\
E$^{pp}_{HFP}$          & -6.964  &   -3.650 & -2.293 & -5.719  &  -6.864  \\
\hline
E$^{2qp}_{Hfp}$         & 4.198 & 3.544 & 3.455 & 3.563 & 3.292 \\
E$^{2qp}_{HFp}$         & 4.084 & 3.372 & 3.270 & 3.450 & 3.211 \\
E$^{2qp}_{HFP}$	        & 3.229 & 2.504 & 2.360 & 2.682 & 2.530 \\
E$^{2qp}_{e}$           & 2.947 & 2.268 & 2.137 & 2.220 & 2.092 \\
\hline
S$_{hfp}^{2p}$   	&  22.784 & 20.087   & 16.516	& 12.810    &  10.302 \\
S$_{Hfp}^{2p}$  	&  23.137 & 20.439   & 16.846	& 13.320    &  10.861 \\
S$_{H\tilde{F}p}^{2p}$  &  23.064 & 20.292   & 16.794	& 13.397     &  10.891 \\
S$_{HFp}^{2p}$	        &  23.168 & 20.490   & 16.871	& 13.248    &  10.823 \\
S$_{HFP}^{2p}$	        &  23.184 & 20.537   & 16.907	& 13.170    &  10.795 \\
S$_{exp}^{2p}$          &  21.890 & 19.233   & 15.430	& 12.615    &  10.334 \\
\hline
E$_{so,tk}$          & -746.778 & -767.427  & -784.319  &  -797.093  &  -807.699 \\
E$_{e}$              & -746.803 & -767.447  & -784.373  &  -797.225  &  -807.883 \\
E$^{pp}_{so,tk}$     & -6.740  &   -3.538 & -2.166 & -5.196  &  -6.151  \\
E$^{pp}_{e}$         & -6.275  &   -2.808 & -0.804 & -4.893  &  -6.063  \\
\hline
\end{tabular}
\end{center}
\caption{ Binding energies, proton pairing energies, lowest two-quasiproton
excitation energies and two--protons separation
 energies  (in [MeV]) for spherical $N=50$ nuclei. } 
\label{tab:SN}
\end{table}

In the second section  of Table~\ref{tab:SN} we show the proton pairing 
energies 
$E^{pp}_{app.}$ for the ground state  of the same nuclei. The neutron pairing
is zero since we have a major shell closure for these nuclei. These energies 
are defined by the expression $E^{pp}_{app.}~=~- \frac{1}{2}  
{\displaystyle Tr}  \left ( \Delta {\kappa}^* \right)$. One has to notice that
from the three Coulomb terms only the pairing term gives a  contribution
to $E^{pp}_{app.}$, the Hartree and Fock terms only give an indirect contribution
through the changes introduced in the wave function by their
consideration in the variational equations. We have also included in the
table $E^{pp}_{ref}$, which gives us the pairing energy without any Coulomb
contribution.  The difference between $E^{pp}_{ref}$ and $E^{pp}_{hfp}$ gives
us an indication of the size of the pairing Coulomb energy, which is about 
1 MeV for all nuclei. 
When the direct Coulomb term is included in the HFB calculation, $Hfp$,
we observe a slight decrease ( ranging from $-0.150$ MeV to $-0.670$ MeV)
in the pairing energies with respect to the calculation $hfp$. 
The inclusion of the  Fock Coulomb term, $HFp$,  produces the opposite effect, 
we get a slight increase (up to  1.0 MeV) of the pairing energy, again with
respect to $hfp$.  Finally, treating all the contributions of the 
Coulomb force in a self--consistent way, $HFP$, we observe an important 
increase (in absolute value, a decrease) in the values of the proton pairing energy of up to 
$ 4$ MeV for the nucleus \nuc{90}{Zr}. It is interesting to notice that
though, for this nucleus,  the pairing Coulomb term itself  is about 650 keV
large, when considered selfconsistently it grows up to 4 MeV.
 Concerning the Slater approximation, $H\tilde{F}p$, we find again that
it provides results very close to the $Hfp$ ones. 
We observe that the maximum effect of the three contributions occurs
for $Z=40$ ( semi-magic nucleus). This can be easily seen in 
Fig.~\ref{fig:EPTSN}b where a definition similar to eq.~(\ref{percen})
 has been used for the pairing energy.
 Comparing a) and b), we find a
clear difference: the percentage of error in the pairing energy is much bigger 
than in the binding energy. We obtain that even the $Hfp$ 
approximation gives us a good result for the total energy, while  this 
approximation is not good for the pairing energy. As a matter of fact, an 
important increase ( $\geq 1.5$ MeV for all nuclei) is obtained for the 
pairing energy when we compare $HFP$ with all other approaches considered. 

In the third section of  Table~\ref{tab:SN} we present the lowest 
two-quasiproton energies calculated in different approximations. In this case
we have also included the exact solution, HFB$_e$. In line with the decrease
of the pairing energies just discussed, we observe a decrease of the excitation
energies as more terms are included in the variational procedure. This
decrease is particularly strong for the pairing exchange Coulomb term.
 We may conclude from this discussion that  neglecting  the Fock and 
pairing Coulomb terms in the variational equations does not affect 
significantly the total energy but it does influence the pairing properties of
the wave function. Then, if we are interested in properties of spherical nuclei 
for which pairing correlations play an important role, like excited states or  
transition energies, one must be careful with the Coulomb exchange terms. 
Notice that not only the two-quasiparticle excited states will be affected
but also any calculation based on a HFB solution, like the Random Phase
Approximation, will also experience a lowering in the excitation energies
of the low-lying collective states.  
 We can see in  Table~\ref{tab:SN} that the most important
reduction in the pairing energy is due to the introduction of the
Coulomb pairing term in a self--consistent way, though the contribution of
this term itself is not very important, about $~1$ MeV for the considered nuclei.

\begin{figure}[h]
\begin{center}
\parbox[c]{15cm}{\includegraphics[width=15.0cm]{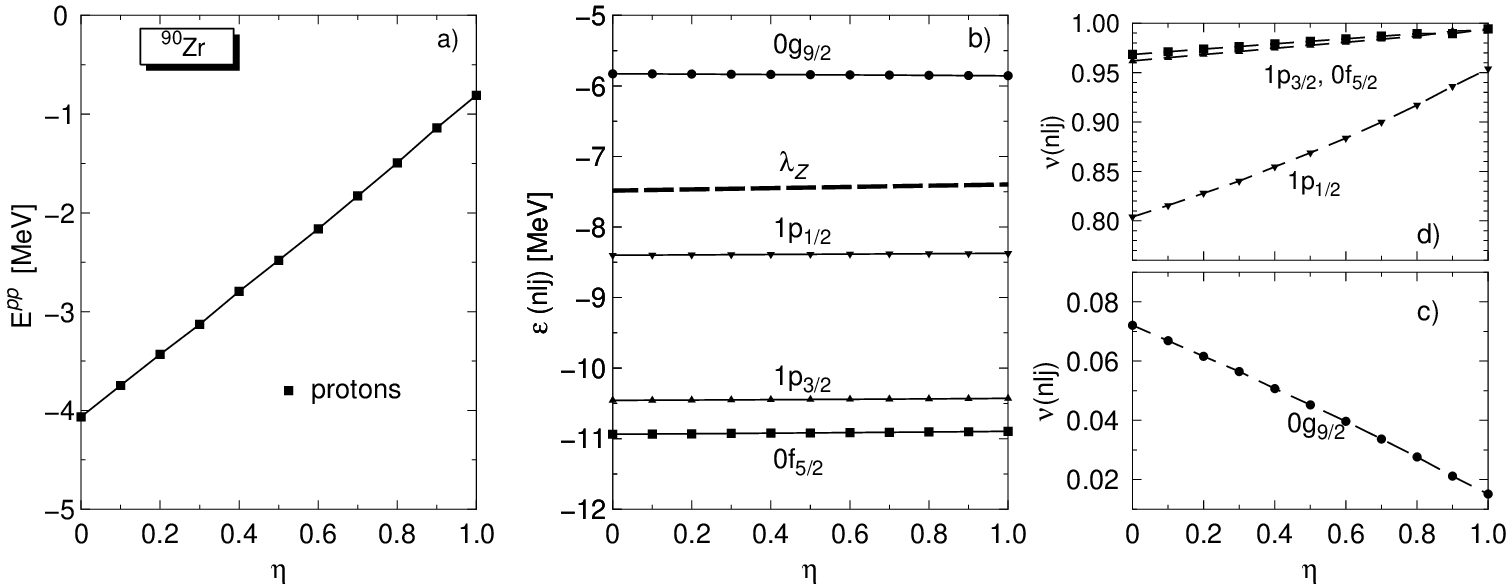}}
\end{center}
\caption{a) E$_{pp}$ versus the $\eta$ parameter for the nucleus \nuc{90}Zr. 
b) Particle energies $\varepsilon$ in the canonical basis and c), d)
occupation numbers $\nu(nlj)$ versus the $\eta$ parameter  
for the nucleus \nuc{90}{Zr}.}
\label{fig:90Zr}
\end{figure}

From the results just discussed one could conclude that the Coulomb interaction
produces an antipairing effect in nuclei. In order to understand this effect
we shall now  analyze in a closer detail the behavior of the most relevant
Coulomb term concerning pairing correlations, namely the pairing Coulomb
term. We have done an additional test for the
nucleus \nuc{90}{Zr}. First of all, an exact HFB calculation (including the
three Coulomb terms and the 
spin--orbit and two--body kinetic mass correction contributions to the pairing
field, i.e., HFB$_e$) has been done for this nucleus. Then, we do a progressive 
{\em switching off} of 
the Coulomb pairing contribution by a parameter $\eta$. For example, $\eta=1.0$
means that we take into account the full Coulomb pairing selfconsistently, 
$\eta=0.6$ means that 60\%  of it is taken selfconsistently and 40\% 
perturbationally and so on. 
In fig.~\ref{fig:90Zr}a we display the proton--proton correlation energy versus 
the $\eta$ parameter for the nucleus \nuc{90}{Zr}. 
We find  a linear behavior, which means that the change in the pairing energy 
due to the {\em switching off} of the Coulomb pairing contribution is produced 
in a continuous way. The solution
that we obtain for $\eta=1.0$ is very similar to the solution for $\eta=0$ 
except for a decrease in the proton--proton pairing energy.

In fig.~\ref{fig:90Zr}b) we present the proton single--particle energies, 
$\varepsilon{(nlj)}$, outside the Z$=28$ shell closure in the canonical basis, 
and in fig.~\ref{fig:90Zr}c),d) the occupation $\nu(nlj)$ for each level $(nlj)$,
 which is given by
\begin{equation}
\nu(nlj) = \frac{1}{2j+1 }
\sum_m \langle \Phi | c_{nljm}^{\dagger} c_{nljm} | \Phi \rangle
\end{equation}

In this figure we observe that the single--particle energies $\varepsilon$ 
barely change with $\eta$. With respect to the occupation numbers, we observe 
that the
Coulomb pairing contribution goes in the direction to fill up the levels 
$0f_{5/2}$, $1p_{3/2}$ and $1p_{1/2}$ and to empty the level $0g_{9/2}$. 
That means, the Coulomb pairing term tries to increase the occupancy
of the levels below the Fermi level at the cost of decreasing the occupancy
of the levels above.

In  the forth section of Table~\ref{tab:SN} we show the two--proton separation 
energies $S^{2p}(N,Z) = E(N,Z-2) - E(N,Z)$.  As we can see in the
table we obtain similar results in all approximations and  the agreement with the 
experiment is good in all cases.  We can understand this result if we consider
that the binding energies are well described in all approximations and that
that $S^{2p}$ is not an observable related with the pairing properties of the
nuclei, which are really sensitive to the different approximations.

Finally, to complete the study of the neglected terms we shall now investigate
the contributions of the spin-orbit term and of the
two-body correction of the kinetic energy  to the pairing field. In the fifth
 section of 
Table~\ref{tab:SN} we show the binding energies $E_{so,tk} $ obtained by 
adding these terms at the end of the minimization of eq.~(\ref{EHFP}), i.e.,  
\begin{equation}
 E_{so,tk} =  E_{HFP} + [V^P_{SO} + V^P_{TK}] \label{Esothk}
\end{equation}
in a similar notation as introduced above.  These terms are repulsive and
a few hundred keV large. If these two terms are now taken into account 
selfconsistently we obtain the exact solution, i.e., HFB$_e$, of the full
 HFB equations,  eq.~(\ref{hfbe}). The resulting energies are on the average 
100 keV  smaller than if
the terms are considered perturbationally.
The last two entries of the table are the pairing energies $E^{pp}_{so,tk}$, 
defined as before. We observe that these terms provide an additional
quenching of the pairing correlations. The behavior is similar to the
one observed with the other terms, i.e., the largest quenching 
takes place for the semi-magical nucleus \nuc{90}{Zr}. 
For this particular nucleus we have calculated the effect of taking into
account selfconsistently only one term and the other one perturbationally, i.e.,
we have performed the following calculations
\begin{equation}
 E_{SO,tk}  =  E_{HFP} + V^P_{SO} + [V^P_{TK}] 
\end{equation}
\begin{equation}
 E_{so,TK}  =  E_{HFP} + V^P_{TK} + [V^P_{SO} ]. 
\end{equation}
We obtain $E^{pp}_{SO,tk}= -1.671$ MeV, $E_{SO,tk}= -784.360$ MeV and
$E^{pp}_{so,TK}= -1.355$ MeV, $E_{so,TK}= -784.354$ MeV, here again capital
letters mean selfconsistent and small letters perturbational approaches. As we
can see each term provides about half of the total effect.

\subsection{The N$=82$ and  N$=126$ regions}
\label{sec:SN82}

In this section we focus on the chains of the N$=82$  and N$=126$ isotones, 
again with neutron shell closure. Then, only proton pairing correlations are 
going to be important in this region.  
Two approximations are used in this case: the approximation
HFB$_s$, see eq.~(\ref{hfba}), which neglects all the terms mentioned in the 
introduction, and the exact calculation HFB$_e$, see eq.~(\ref{hfbe}), which 
takes into account all the contributions  from the different terms of 
the force to the fields. In the HFB$_s$ approximation, the missing 
terms are not added at the end of the variational procedure. 
In order to be able to compare the total binding energies in the HFB$_s$
 approach with the exact ones one must add these terms perturbativaly,
 i.e, the HFB$_{s+}$ approximation.
 In the N$=82$ region, the basis is determined by $q=1.0$, $p=1.0$ and 
$N_0=11.1$ and in this case the parameterization D1 is used. In the N$=126$ 
region, the basis $q=1.0$, $p=1.0$ and $N_0=12.1$ has been used and for the 
force  parameterization we have chosen  D1S.

In the first two columns of Table~\ref{tab:N82} we display the total binding 
energies. We find that the approach HFB$_{s+}$ provides a very good approximation 
to the exact results. In the last three columns of the same table we show the
 proton--proton pairing energies obtained in the approximations HFB$_s$, 
HFB$_{s+}$  and the exact ones HFB$_e$.   
Both approximations provide more proton pairing correlations
that the exact calculation HFB$_e$. We can see that,  contrary to the 
total binding energy case, the perturbational calculation HFB$_{s+}$ is not
a good approximation for the pairing energy. We do not find any noticeable
difference between both mass regions or the Gogny parameterizations.

From these results  we may conclude that, for spherical nuclei, the
approximation of ignoring some terms of  the hamiltonian, HFB$_s$, presents
some differences with the exact calculation.
The approximation of considering these terms at least
in first order perturbation theory, HFB$_{s+}$,   works
well for observables related to, or that can be obtained from, total binding
energies. On the other hand, the wave function content may be different
from the one obtained in the exact calculation.

\begin{table}[h]
\begin{center}
\begin{tabular}{lccccc}
\hline 
Nucleus            &  E$_{s+}$  &   E$_e$    &E$^{pp}_s$&E$^{pp}_{s+}$&E$^{pp}_e$  \\
\hline
\nucZ{50}{132}{Sn} & -1090.875  & -1090.891 & -0.0000	& -0.0000  & -0.0000   \\
\nucZ{52}{134}{Te} & -1111.079  & -1111.236 & -7.4700	& -6.5881  & -4.7615   \\
\nucZ{54}{136}{Xe} & -1129.549  & -1129.784 & -11.7377  & -10.3805 & -7.4571   \\
\nucZ{56}{138}{Ba} & -1146.233  & -1146.557 & -14.3629  & -12.7469 & -8.6469   \\
\nucZ{58}{140}{Ce} & -1161.102  & -1161.511 & -15.7718  & -14.0409 & -8.8442   \\
\hline
\nucZ{80}{206}{Hg} & -1615.860  & -1616.038 &  -6.6570  & -5.8068  & -4.2799   \\
\nucZ{82}{208}{Pb} & -1634.620  & -1634.639 &  -0.0000  & -0.0000  & -0.0000   \\
\nucZ{84}{210}{Po} & -1642.746  & -1642.885 &  -6.1489  & -5.3940  & -3.9841   \\
\nucZ{86}{212}{Rn} & -1649.656  & -1649.878 & -10.5256  & -9.2758  & -6.6871   \\
\nucZ{88}{214}{Ra} & -1655.273  & -1655.560 & -13.7754  & -12.1942 & -8.2135   \\

\hline		   	        	  	     		  	      
\end{tabular}
\end{center}
\caption{ Binding energies  and proton pairing energies (in [MeV]) for spherical
 $N=82$ and $N=126$ nuclei with the methods HFB$_s$ and 
HFB$_e$ for $I=0 \hbar $.} 
\label{tab:N82}
\end{table}

\section{Normal deformed nuclei}
  In this section we will study the effects  neglecting  some terms
in the hamiltonian for deformed nuclei. In particular we shall investigate 
their behavior at high angular momentum as well as the impact of the exchange
terms on the neutron system.

\subsection{Rare earth nuclei}
\label{sec:REN}

In this region, we study  six nuclei: the three Erbium isotopes  
 \nuc{164,166,168}{Er}, and the three Ytterbium
isotopes, \nuc{162,164,166}{Yb}. 
As in the preceding subsection we shall investigate only the approximation
HFB$_s$ (or HFB$_{s+}$ when the last one does not make sense)
 to compare with the exact calculation HFB$_e$ and
the experimental results when available.
 For these calculations we use the basis with q$=1.3$, p$=1.0$ and 
N$_0=11.1$ and the D1S parameterization of the force.

\begin{table}[h]
\begin{center}
\begin{tabular}{lcccccc}
& \nuc{164}{Er} & \nuc{166}{Er} &  \nuc{168}{Er} & \nuc{162}{Yb} & 
  \nuc{164}{Yb} & \nuc{166}{Yb} \\
\hline
E$_{s+}$ & -1328.665 & -1343.424 & -1357.391 &  -1305.810 & -1322.764 & -1339.332 \\
E$_{e}$  & -1329.176 & -1343.868 & -1357.777 &  -1306.361 & -1323.425 & -1340.123 \\

\hline
E$^{pp,\pi}_s$ & -7.068  & -6.297  &  -5.642 & -10.081 & -9.354  & -8.309 \\
E$^{pp,\pi}_e$ & -1.723  & -1.727  &  -1.730 & -4.167  & -0.001 & -0.000 \\
E$^{pp,\nu}_s$ & -8.757  & -7.475  &  -5.602 & -9.444  & -9.449  & -9.049  \\
E$^{pp,\nu}_e$ & -7.756  & -6.310  &  -4.310 & -8.223  & -11.005 & -9.614 \\
\hline
Q$_s$          &  7.590 & 7.781  &  7.893 &  6.073 & 6.871  & 7.563 \\
Q$_e$          &  7.552 & 7.712  &  7.836 &  6.190 & 7.602  & 7.787 \\
\hline
\end{tabular}
\end{center}
\caption{Ground state binding energies, pairing energies (in [MeV]), 
and quadrupole momenta (in barns), and in the approximation (HFB$_s$)
 and in the exact one (HFB$_e$). }
\label{tab:RE-EpQ0}
\end{table}

We shall first investigate the ground state properties of these nuclei.
In the upper part of  Table~\ref{tab:RE-EpQ0} we show the binding energies.
We find that the approximate calculation, HFB$_{s+}$, provides results very 
close to the
exact ones. In the middle of the table we display the pairing energies
calculated in both approaches.     
We find that the exact calculation provides proton pairing energies much
smaller, in absolute value,  than the approximate calculation, HFB$_s$.
 The global effect that we observe in the HFB$_e$ calculations is a decrease 
 of the absolute value of the proton pairing correlations 
compared to the approximate calculation. For 
\nuc{164}{Yb} and \nuc{166}{Yb}  these correlations even  vanish.
 The behavior of the neutron pairing 
energy is different depending on the nucleus: for the last two nuclei 
$|E^{pp,\nu}_e|$  is increased by about $1$ MeV  and for the other
nuclei it is decreased by approximately  the same amount as compared to
$|E^{pp,\nu}_s|$.
In general, we can say that the global effect of including the 
Coulomb exchange and pairing contributions in the self--consistent procedure 
is a strong reduction of the absolute value of the proton pairing correlations 
in all the cases.  
The circumstance that the inclusion of the neglected terms influences more
the proton pairing than the neutron one is obviously due to the Coulomb
terms that do not affect the neutron system.
With respect to the quadrupole moments we observe that the relative 
variation between both calculations is relative small, approximately $0.5\%$ 
for all the nuclei except for \nuc{164}{Yb} where Q increases in 
the exact calculation by about $10\%$.

 The fact that the total binding energies are very close in both approaches  
 is the reason why the non self--consistent and perturbational 
approximation has been considered in the past as a good approximation 
in spite of the fact that the wave functions are different.  
\begin{figure}[h]
\begin{center}
\parbox[c]{14cm}{\includegraphics[width=14.0cm]{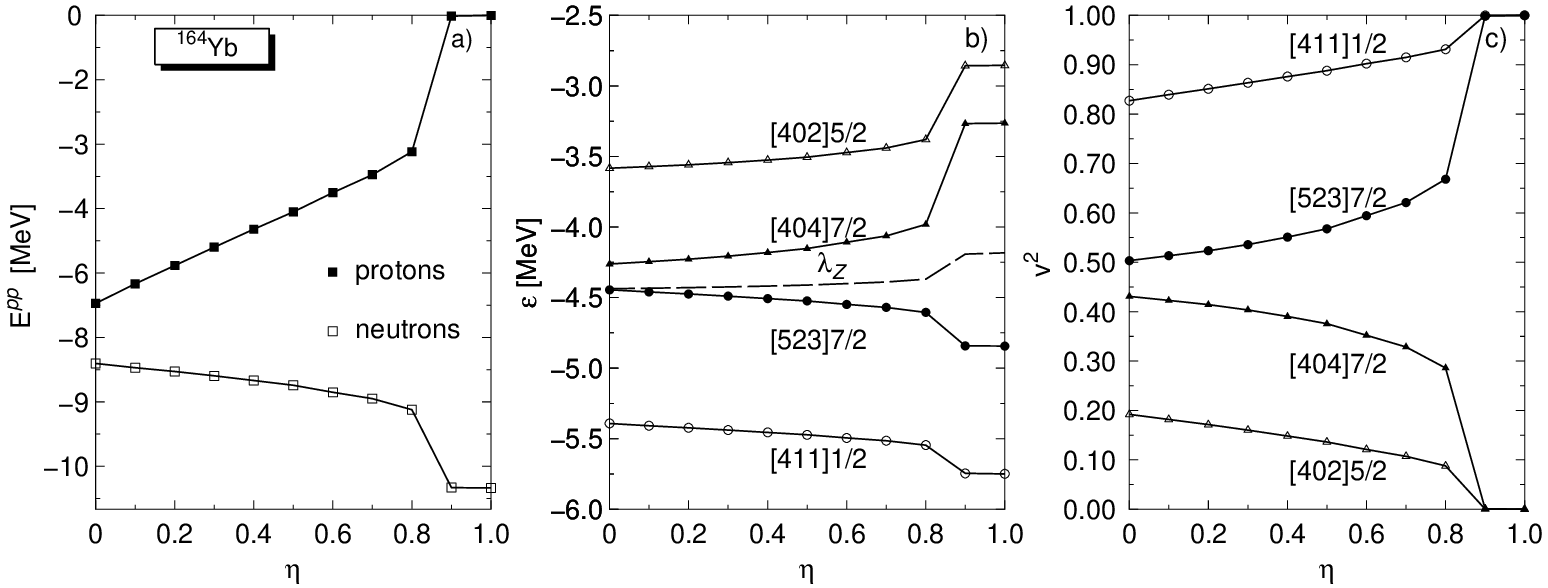}}
\end{center}
\caption{a) Variation of the pairing energies versus the $\eta$ parameter for the
nucleus \nuc{164}{Yb}. 
b) Proton single particle energies $\varepsilon$ (in MeV). The Fermi level is 
the dashed line.
c) Occupations $v^2$ of the levels in the canonical basis, versus the
$\eta$ parameter, for the nucleus \nuc{164}{Yb}. The  levels have been labeled
by the quantum numbers $[n\,n_z\,l_z]\Omega$ in an axial basis.}
\label{fig:164Yb}
\end{figure}

\begin{figure}[h]
\begin{center}
\parbox[c]{12cm}{\includegraphics[width=12cm]{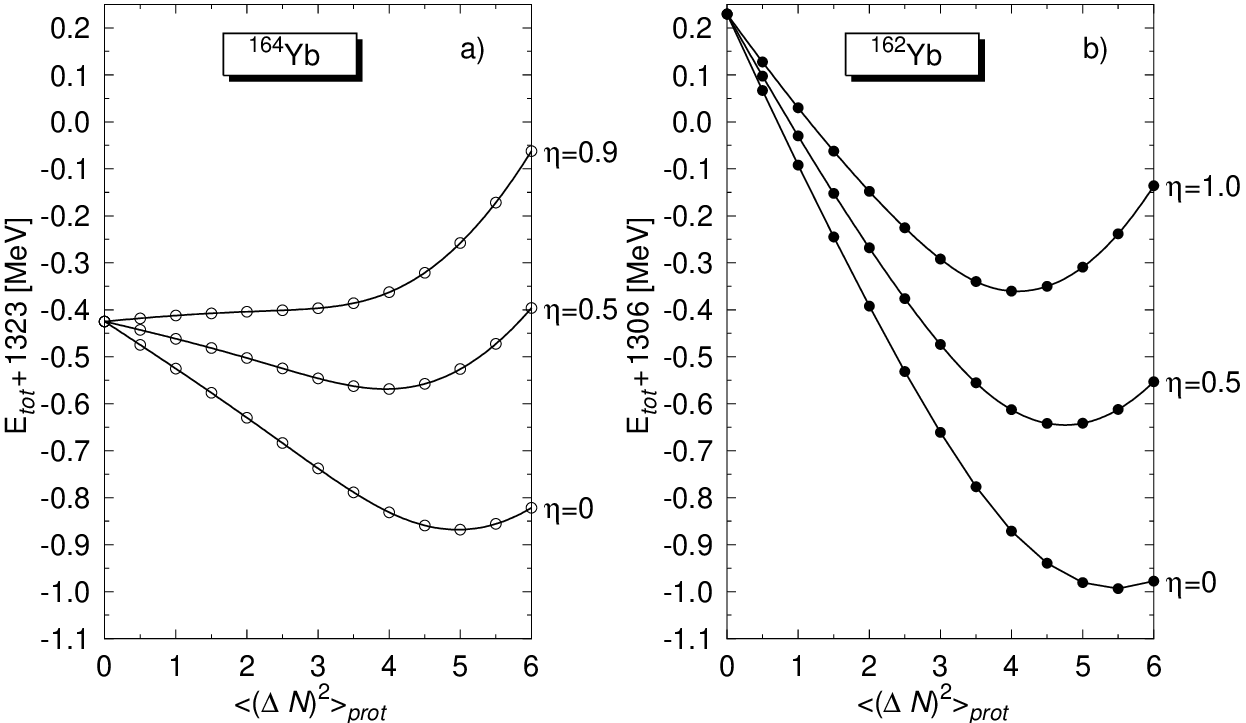}}
\end{center}
\caption{Binding energy versus the constrained proton fluctuation for
different $\eta$ values.} 
\label{fig:dn2}
\end{figure}

The proton pairing energies of the nuclei \nuc{164}{Yb} and \nuc{166}{Yb}
show an anomalous behavior: For these nuclei, the proton pairing   correlations 
 vanish completely in the HFB$_e$ approach. Taking into account the behavior of 
the Er isotopes in Table~\ref{tab:RE-EpQ0}, it would be expected to obtain for 
the nuclei \nuc{164}{Yb} and \nuc{166}{Yb} approximately $-4$ MeV for the 
proton pairing energy. In order to understand this quenching we have performed
a calculation  for  \nuc{164}{Yb} similar to the one done for the nucleus 
\nuc{90}{Zr} switching off progressively the Coulomb pairing contribution.
  In Fig.~\ref{fig:164Yb}a) the proton and neutron pairing energies 
are displayed versus the $\eta$ parameter. As it can be seen from this
figure, increasing the $\eta$ value up to $\eta=0.8$ produces a linear 
increase of the proton pairing energy. However, for $\eta=0.9$ a sharp 
increase is obtained, going to the unpaired solution.
By extrapolation of the small $\eta$ values to $\eta=1.0$ the expected proton
pairing energy for this value would be around $-2.5$ MeV.
However, the selfconsistent solution at $\eta=1.0$ corresponds to a non-paired
proton system, which seems to indicate that the mean field approximation breaks 
down collapsing to the non-correlated regime. 
In Fig.~\ref{fig:164Yb}b) we show the proton single particle energies in the 
canonical
basis versus the $\eta$ parameter. Here we can observe the effect of the sharp 
phase transition on the single particle energies. In Fig.~\ref{fig:164Yb}c) we 
can see how the occupations of the levels above (below) the Fermi level get 
empty (full). This sharp phase transition will explain the anomalous
behavior of these two nuclei. The phase transition itself has more to do with
the mean field approach than with the exchange terms. This can be seen
very clearly in Fig.\ref{fig:dn2}a) where we show potential energy curves as
a function of the proton fluctuation for different $\eta$ values. That means
we have solved the HFB equations with an additional constraint on the w.f., 
namely the fluctuation in the proton number. For \nuc{164}{Yb} and $\eta = 0$
we find a well defined minimum around the superfluid solution, 
$\langle (\Delta N)^2 \rangle_\pi = 5$, i.e., the mean field approach makes
sense. For $\eta =  0.9$, however, the energy surface is quite
flat indicating that a theory able to include further correlations would be
more appropriate.  For \nuc{162}{Yb}, where no pairing collapse takes place,
we find well defined minima for all $\eta $ values, see  fig.~\ref{fig:dn2}b).

\begin{table}
\begin{center}
\begin{tabular}{cccccc}
{\sc Nucleus} & $E_{\gamma}$ (EXP) & $E_{\gamma}$ (HFB$_s$) & $E_{\gamma}$ (HFB$_s+$) &
$E_{\gamma}$ (HFB$_e$) & $\Delta E_{\gamma}(\%)$ \\
\hline
\nuc{164}{Er} & 91.4  &  82.8  & 67.7  &  68.4  &  21.1  \\
\nuc{166}{Er} & 80.6  &  77.6  & 63.5  &  65.0  &  19.4 \\
\nuc{168}{Er} & 79.8  &  73.0  & 59.4  &  60.7  &  20.3 \\
\nuc{162}{Yb} & 166.9 & 115.7  & 90.3  &  93.9  &  23.2 \\
\nuc{164}{Yb} & 123.4 &  98.7  & 85.5  &  80.8  &  22.2 \\
\nuc{166}{Yb} & 102.4 &  88.8  & 75.9  &  75.8  &  17.2 \\
\end{tabular}
\end{center}
\caption{Experimental (\nuc{162}{Yb} \cite{ND162}, \nuc{164}{Er} and 
\nuc{164}{Yb} \cite{ND164}, \nuc{166}{Er} and \nuc{166}{Yb} \cite{ND166} and
\nuc{168}{Er} \cite{ND168}) and theoretical transition energies from $I=2\hbar$ 
to $I=0\hbar$ obtained with the methods HFB$_s$, the perturbational one 
HFB$_{s+}$, 
the exact calculation HFB$_e$ (in keV) and the relative variation between 
HFB$_s$ and HFB$_e$ ($\%$).}
\label{tab:RE-Egam}
\end{table}

\begin{figure}[h]
\begin{center}
\parbox[c]{12cm}{\includegraphics[width=12cm]{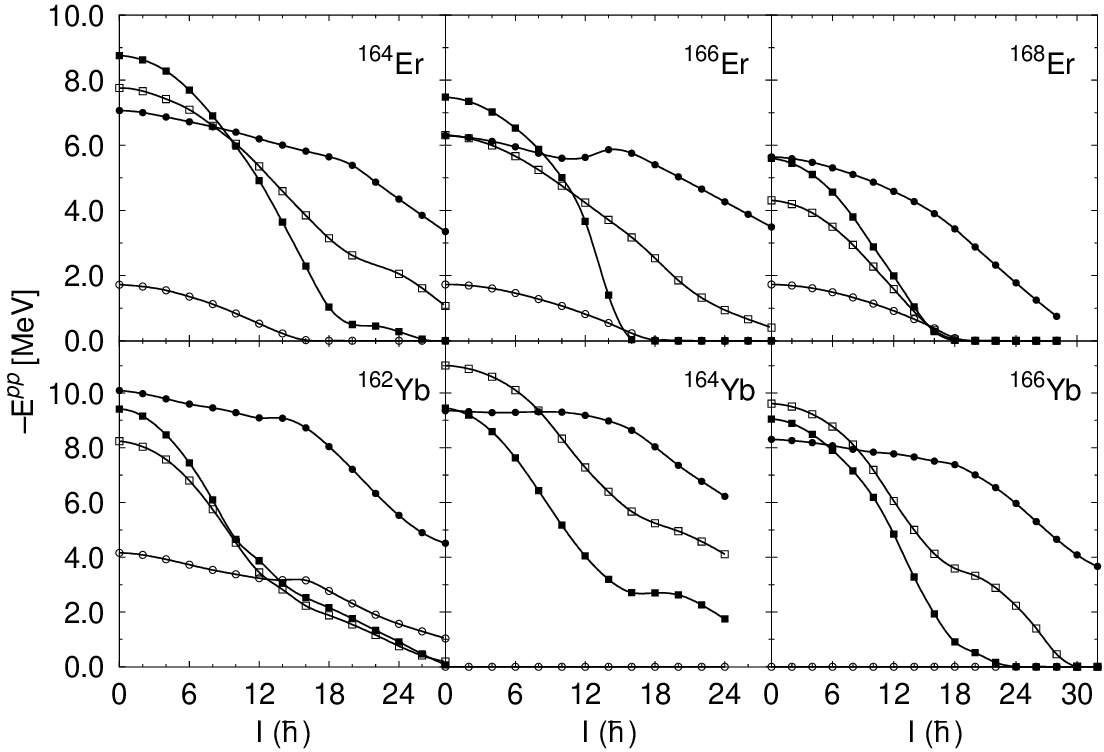}}
\end{center}
\caption{E$^{pp}$[MeV] versus angular momentum calculated with 
HFB$_s$ (full symbols), HFB$_e$ (empty symbols). The circles (boxes ) are
for protons (neutrons).} 
\label{fig:RE-EP}
\end{figure}

  In order to investigate the  high spin properties of these nuclei we have 
solved the cranking HFB equations, eq.~(\ref{chfb}), in both approximations 
for the same nuclei.
In Fig.~\ref{fig:RE-EP}, we display the pairing energies, $-E^{pp}$, as a 
function of the angular momentum. We find large differences between the
approximate and the exact results. The absolute values of the 
HFB$_s$ proton pairing energies are 
much larger than the exact ones, the difference being not only quantitative
but also qualitative for the nuclei \nuc{164}{Yb} and \nuc{166}{Yb}. These
nuclei
remain super-fluid for the whole spin range in the approximate solution
and normal fluid in the exact one. For the neutron system, the  behavior
is qualitatively similar in both approaches, the small differences observed
depend on the nucleus considered. 
 The strong quenching of the pairing energies obtained in the exact calculations
will affect some observables, for example, the transition energies  along 
the yrast band. In  Table~\ref{tab:RE-Egam} we show the 
E$_{\gamma}(2^+ \longrightarrow 0^+)$ values obtained in the approximations
HFB$_{s}$, HFB$_{s+}$ and in the exact calculation.
As it can be observed, to consider the exchange Coulomb term and all the 
contributions to the pairing field leads to a decrease in the 
transition energies, with a relative change around $20\%$. Since the 
perturbational calculation up to first order (HFB$_{s+}$) provides a good
approximation to the total binding energies we expect that the 
transition energies calculated in this way will also be close to the exact 
ones. As we can see in the table this is the case. 
On the other hand, by comparing our results with the experimental
ones, which are included in the same table, we  observe that
the HFB$_e$ method gives a worse agreement with the experiment than the
HFB$_{s+}$ one.

It is interesting to know what happens with the   D1 parameterization
of the Gogny force. We have done the calculation with this parameterization 
for the nucleus \nuc{164}{Er}, we obtain the value 
$E_{\gamma}= 95.7$ keV with HFB$_s$ and $E_{\gamma}=79.0$ keV with HFB$_e$, 
with a relative variation between both calculations of $21.1\,\%$, 
the same relative variation as obtained with the D1S parameterization, as we can 
see in the Table~\ref{tab:RE-Egam}. The two values of the transition energy 
E$_{\gamma}$ obtained with the D1 are closer to the experiment, but again the 
approximate calculation HFB$_s$ gives  the best agreement with the 
experiment. 

\begin{figure}[h]
\begin{center}
\parbox[c]{14cm}{\includegraphics[width=14cm]{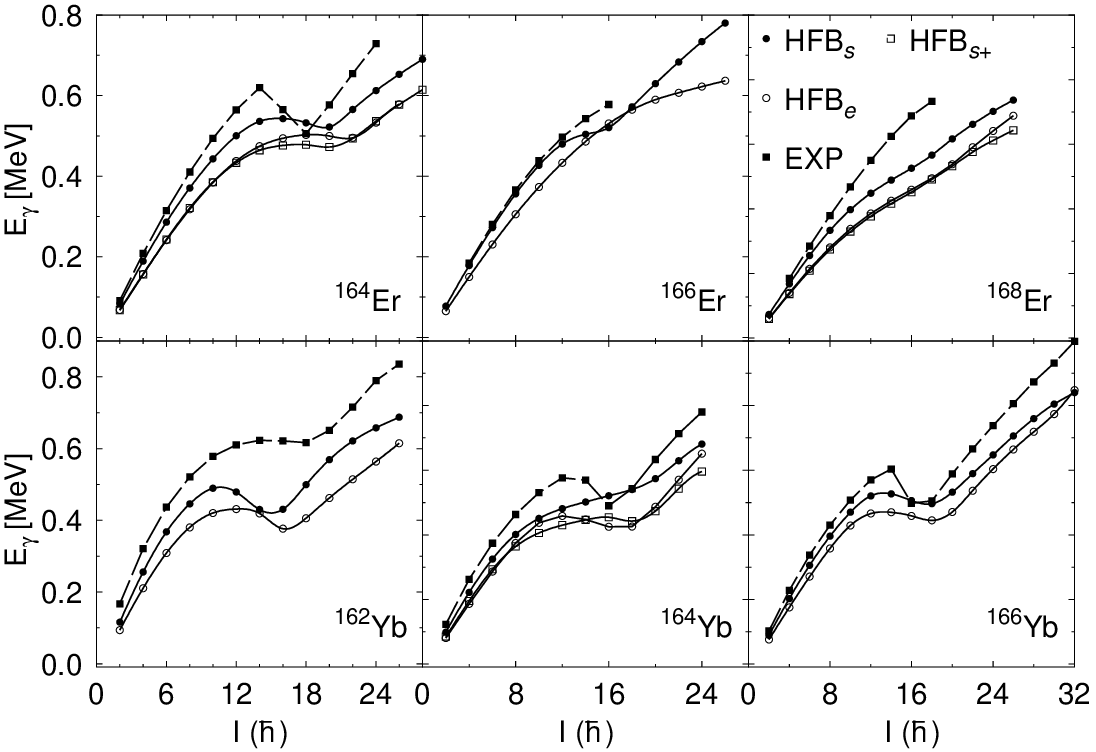}}
\end{center}
\caption{E$_{\gamma}$[MeV] versus angular momentum calculated with 
HFB$_s$ (full circles), HFB$_e$ (empty) and the experimental values 
(full squares). The empty squares in the panels of the nuclei 
\nuc{164}{Er}, \nuc{168}{Er} and \nuc{164}{Yb} represent the HFB$_{s+}$
approach.
 The experimental values are from: \nuc{162}{Yb} \cite{ND162}, 
\nuc{164}{Er} and  \nuc{164}{Yb} \cite{ND164}, \nuc{166}{Er} and \nuc{166}{Yb} 
\cite{ND166} and \nuc{168}{Er} \cite{ND168}.}
\label{fig:RE-Eg}
\end{figure}

The behavior of the gamma ray energies at high spin is displayed in 
Fig.~\ref{fig:RE-Eg}  for the six analyzed nuclei, together with the 
experimental data. We find that the results of the approximation HFB$_s$
are rather different from the exact ones as one expects from the disparity 
of the pairing energies found in the two calculations. For the nuclei 
\nuc{164}{Er}, \nuc{168}{Er} and \nuc{164}{Yb} we have also included the 
approximation HFB$_{s+}$. As expected the agreement with the exact calculation
is much better than the one obtained in the HFB$_{s}$ calculation.
We have also checked the Slater approximation for deformed nuclei and at
high angular momentum. That means, we have evaluated the Fock exchange 
Coulomb term in the Slater approximation and added it to the energy $E_s$
of eq.~(\ref{hfba}) {\it before} the variation. The results of these
calculations, quadrupole moments, pairing energies and gamma ray energies 
along the Yrast band, practically coincide with the results of the plain  
HFB$_s$ calculation.

\begin{figure}[h]
\begin{center}
\parbox[c]{12cm}{\includegraphics[width=12cm]{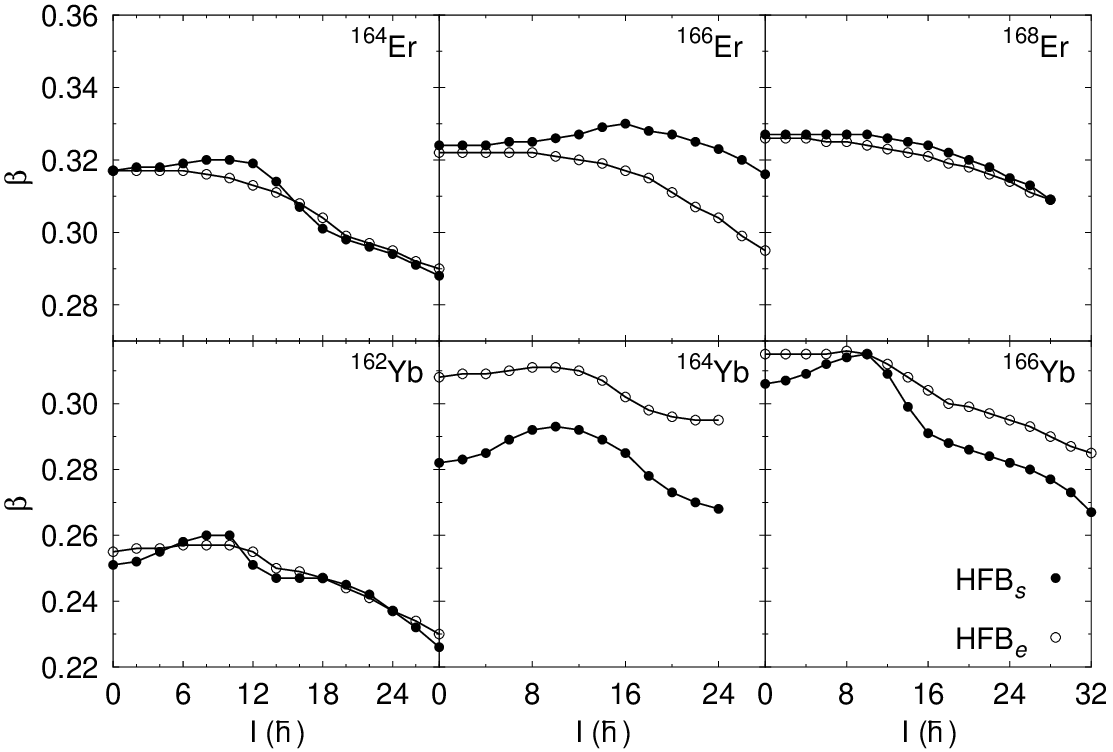}}
\end{center}
\caption{The $\beta$ deformation parameter versus the angular momentum 
calculated with HFB$_s$ (full symbols), HFB$_e$ (empty symbols).} 
\label{betas}
\end{figure}
From the results discussed up to now, it seems that it is enough to consider
the neglected terms in first order perturbation theory. This, however is not
quite true since some observables can not be expressed as a function of the 
binding energies. For instance, the transition probabilities, magnetic moments, 
nuclear radii, etc.  It is interesting to investigate the predictions for 
some of these
observables in the exact and in the approximated calculation.
In Fig.~\ref{betas}, we show the beta deformations as a function of the
angular momentum for the six nuclei under study. The beta deformation parameter
is mainly affected by the alignment effects and the pairing correlations. In
general, larger  pairing correlations imply smaller deformations and 
larger alignments smaller  $\beta$-deformations (Coriolis anti-stretching effect).  
 In the exact calculations, in general, we have smaller absolute values of the
 pairing correlations
and larger alignment. In principle, one could expect a compensation and only
in cases where one of the effects is much larger than the other one we should
obtain a change in deformation  corresponding to the largest effect. 
In the figure
we observe that in the nuclei  \nuc{164}{Er}, \nuc{168}{Er} and \nuc{162}{Yb} 
such a
compensation takes place. In \nuc{166}{Er} the compensation takes place for small 
spins but not for large ones and in  \nuc{164}{Yb} and \nuc{166}{Yb} there
is no compensation at all. Looking in Fig.~\ref{fig:RE-Eg}, we find that in 
\nuc{166}{Er}, there is clearly a stronger alignment in the exact solution
than in the approximated one, which causes an smaller deformation in the exact
calculation at high spins. In \nuc{164}{Yb} and \nuc{166}{Yb} the large
quenching observed in the pairing energy explains the distinct behavior of the
exact solution from the approximate one.
\begin{figure}[h]
\begin{center}
\parbox[c]{12cm}{\includegraphics[width=12cm]{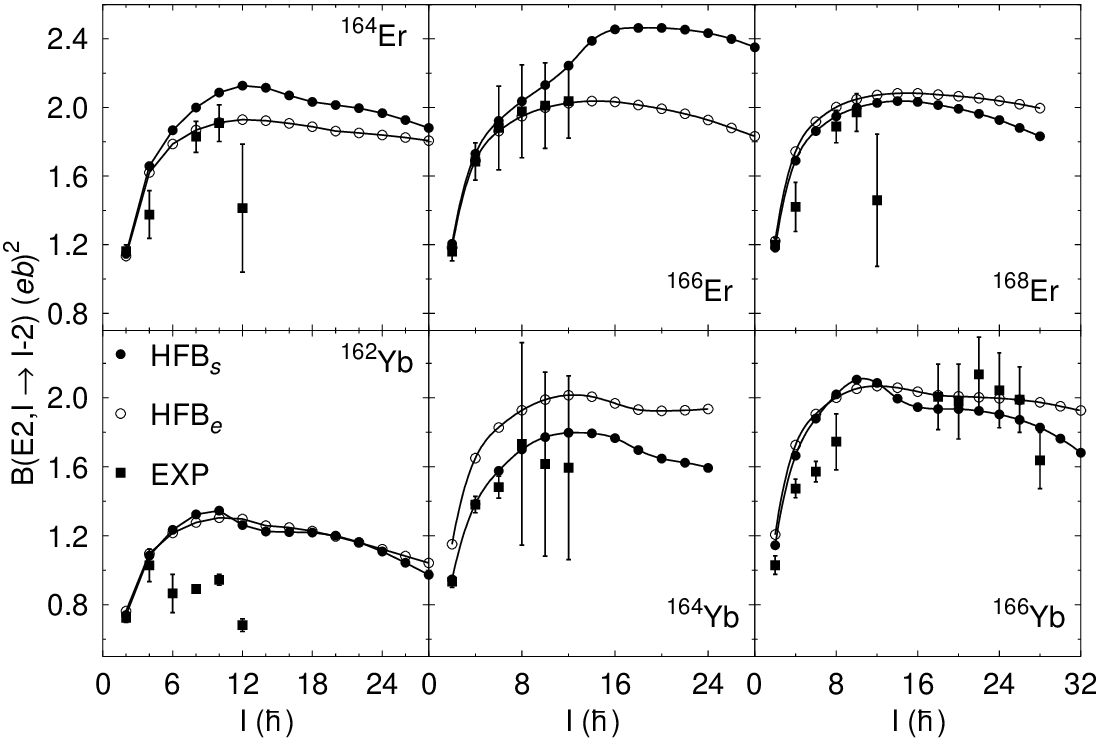}}
\end{center}
\caption{ Reduced transition probabilities versus angular momentum calculated 
with HFB$_s$ (full circles), HFB$_e$ (empty) and the experimental ones
(full squares). The experimental values are from \nuc{162}{Yb} \cite{ND162}, 
\nuc{164}{Er} and \nuc{164}{Yb} \cite{ND164}, \nuc{166}{Er} and \nuc{166}{Yb} 
\cite{ND166} and \nuc{168}{Er} \cite{ND168}.}
\label{fig:REbe2}
\end{figure}
In Fig.~\ref{fig:REbe2} we present the reduced transition probabilities
B(E2) for the nuclei we are analyzing. 
 Though these probabilities are related with the $\beta$-deformations we 
expect larger differences between the two calculations in the  B(E2) case 
because this quantity has
to do only with protons while to the $\beta$-deformations both the 
proton and the neutron systems  contribute. We should remember that 
neglecting  the
Coulomb exchange terms affects only the proton system. We find this 
supposition to be right, specially at high spins. For the nucleus 
\nuc{164}{Yb} there are differences even at zero angular momentum due to the 
change of deformation in the ground state (see Table~\ref{tab:RE-EpQ0}).
Though we find some differences between both calculation, in the 
comparison with the experimental results it does not matter which one we take.

\begin{figure}[h]
\begin{center}
\parbox[c]{12cm}{\includegraphics[width=12cm]{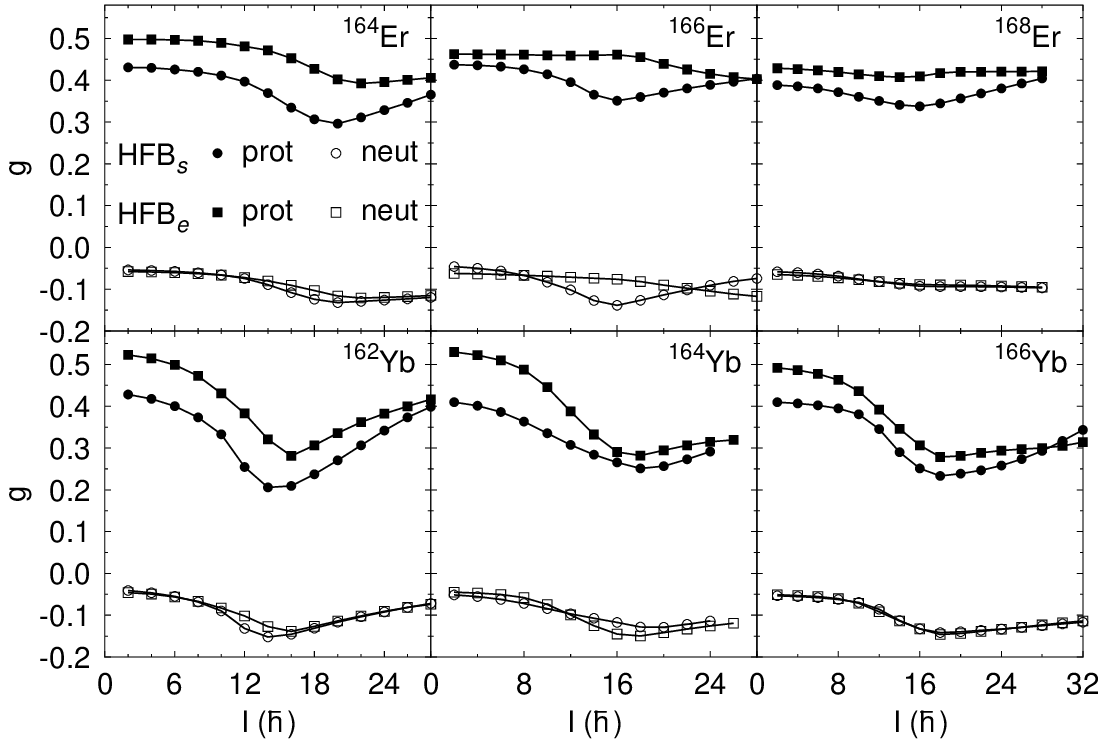}}
\end{center}
\caption{Gyromagnetic factors versus angular momentum calculated with 
HFB$_s$ (circles), HFB$_e$ (squares). Full (empty) symbols are for 
protons (neutrons).}
\label{fig:gpgn}
\end{figure}

In Fig.~\ref{fig:gpgn} we finally present the gyromagnetic factors versus
the angular momentum. For this observable we expect only small differences 
between the two
calculations for the neutron gyromagnetic factor, $g_n$, for the same reasons as
before, and somewhat larger for the proton gyromagnetic factor $g_p$. We find
this supposition to be right. In the HFB$_s$ approximation,
the general behavior of these nuclei is, first, a more or less sharp neutron
(i$_{13/2}$) alignment take place at medium spin values and then a
proton (h$_{11/2}$) alignment take place at high spins. In the HFB$_e$
calculation the situation is qualitatively similar with the exception of 
the nucleus \nuc{166}{Er} which shows a very delayed neutron alignment, contrary
to the experimental situation \cite{ND166}.

\subsection{Actinide region: The nucleus \nuc{240}{Pu}.}
\label{sec:AR}
\begin{figure}[h]
\begin{center}
\parbox[c]{14.cm}{\includegraphics[width=14.cm]{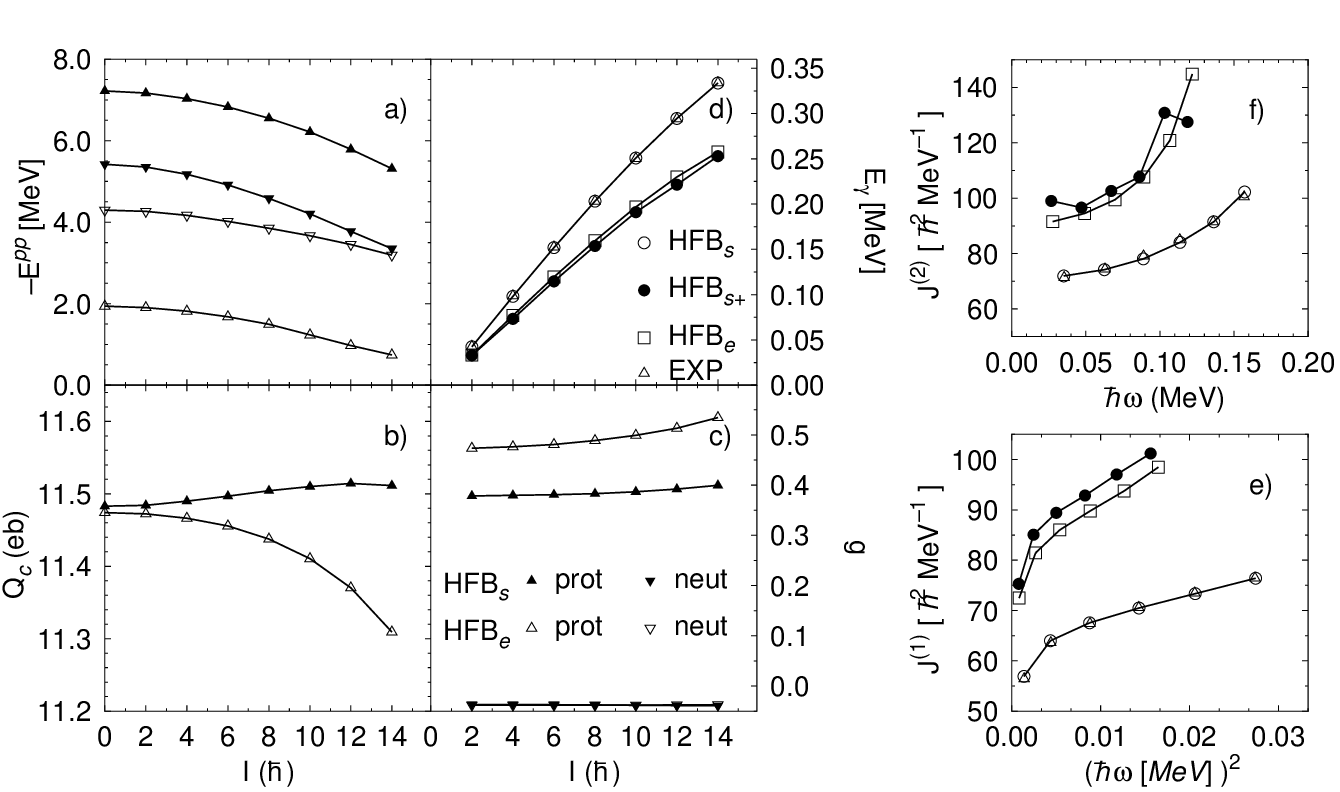}}
\end{center}
\caption{ Properties of the nucleus \nuc{240}{Pu} along the {\em yrast} band 
in  the HFB$_s$ and HFB$_e$ approaches.
a) Pairing energies.   b) Charge quadrupole moments.
c)Gyromagnetic factors. d) Gamma ray energies. 
 e) Moments of inertia ${\mathcal J}^{1}$ and f): 
Moments of inertia ${\mathcal J}^{(2)}$. The labels for panels a), b) and c)
are the ones of panel c).  The labels for d), e) and f) are listed in d).  
 The experimental data are from \cite{ND240}.}
\label{fig:Pu}
\end{figure}

 In this subsection we shall now discuss an example of a heavy deformed
nucleus, to see if there is a possible mass dependence of the terms neglected 
in the approximate calculations. As a representative of the actinide region we
 have chosen  the nucleus \nuc{240}{Pu}. In the calculations we have
used the basis  q$=1.3$, p$=1.0$ and N$_0=14.0$ which is large enough 
to guarantee the convergence, and the D1S parameterization of the force. 
Let us first discuss the wave function content of the HFB$_{s}$  approximation
and of the exact theory  HFB$_{e}$.  In Fig.~\ref{fig:Pu}, panel (a), we show 
the pairing
energies $E^{pp}$ in both theories as a function of the angular momentum.  
As before, we find a strong quenching
for the values of the proton system in the HFB$_{e}$ as compared to the 
HFB$_{s}$ one. The values for the neutron system, on the contrary, are very
similar in both  approaches. In panel (b) we display the electric quadrupole
moment. At small angular momentum both predictions are very close but
at large angular momentum they differ by up to $2\%$ at the largest spin
considered. The fact that the HFB$_{e}$ prediction is smaller can be
understood looking at panel (c) where we show the gyromagnetic factors. 
As expected the $g_n$ are very similar in both approaches for all spin values. 
The $g_p$, in the  HFB$_{e}$ prediction indicates a larger proton alignment
than in the HFB$_s$ one, which will cause the observed anti-stretching effect 
in the HFB$_{e}$ approach. 
  With respect to  observables related
to energy differences, we can consider the approaches  HFB$_{s}$, HFB$_{s+}$
and HFB$_{e}$. In the same figure, we present the transition energies 
$E_\gamma$,  panel d), first moment of inertia ${\mathcal J}^{1}$, panel (e),
and second moment of inertia ${\mathcal J}^{2}$, panel (f), in the three 
approaches, as well as the experimental data.
The agreement of the plain HFB$_{s}$ with the experiment for the three
observables is  spectacular. The exact solution HFB$_{e}$, according to
the less pairing-correlated wave function, see panel (a), provides too large
moments of inertia and too small gamma-ray energies.
For the perturbative  approach HFB$_{s+}$, we obtain  values similar
to the exact calculation HFB$_{e}$, though for the more sensitive quantity
${\mathcal J}^{2}$ some differences are found at the smallest and the largest
angular momenta. 

It is very remarkable  that the   HFB$_{s+}$  binding-
and $\gamma$-ray energies are very close to the HFB$_{e}$ ones. 
 This  fact also happens in the other nuclei calculated so far.
To gain some insight in the reasons for this agreement we have calculated
separately all contributions to the total energy in both approaches. 
That means, with the 
wave function $|\Phi\rangle_s$ determined by minimization of $E_s$, we have
evaluated the Fock and pairing term of the Coulomb force, as well as the
pairing terms stemming from the spin-orbit and the two-body kinetic energy 
terms. We have also separately calculated the mentioned terms with the 
wave function $|\Phi\rangle_e$  determined by minimization of the exact energy 
$E_e$.   In the upper part of Table~\ref{tab:240Pu} we show the energy $E_s$ 
of Eq.~(\ref{hfba}) evaluated with the wave function $|\Phi\rangle_s$ 
(approach $s$) and with $|\Phi\rangle_e$ (approach $e$), as well as the 
terms just mentioned  (we use the notation of Eq.~(\ref{VLterms})) for the
ground state of \nuc{240}{Pu}. The last
column, denoted $E_{tot}$, corresponds to the total energy, $E_{s+}$ in the
$s$ approach and  $E_{e}$ in the $e$ one. As expected the binding energy
is lower in the exact than in the perturbative calculation. On the other
hand, since the wave function $|\Phi\rangle_s$ has been determined by
minimizing  $E_s$, it is also obvious that this quantity is lower
in the $s$ approximation than in the $e$ one.  Concerning the other terms,
each of them separately provides a bigger energy lowering in the $e$ than
in the $s$ appoach. We also observe that the relative largest difference
 between both approaches corresponds to the pairing Coulomb term. 
In the lower part of the same Table we show, as a function of the angular
momentum, the energy differences $\Delta \xi = \xi(I) -\xi(I-2)$, where
$\xi$ stands for each energy entry of the upper part of the Table, i.e., 
 $\Delta \xi$ represents
the contribution of each term to the gamma ray energy from the state 
$I$ to $I-2$. These energy differences have been calculated again in
the approaches $s$ and $e$. Let us first concentrate on the $I=2 \hbar$ 
row. The prediction for the gamma ray energy taking into account only
the terms of the standard approach are almost identical independently
of the fact that we evaluate them with $|\Phi\rangle_s$ or with 
$|\Phi\rangle_e$ and the same happens with the other terms. Obviously, if
one adds all these contributions,  the prediction for the total $\gamma$-ray 
energy with all terms is very similar in both approaches. It is also
interesting to note that the contributions from the terms not considered in
the standard approach are all negative and lead to a more compressed spectrum.
We also observe that the largest contribution by far is the one from the
Coulomb pairing term, though the other contributions are not at all negligible.
At higher angular momentum we observe the same trend as for $I=2 \hbar$,
i.e., the energy differences for a given $I$ are very similar in both 
approaches. It seems therefore that the reason, for the good agreement of the
 gamma ray energies in both approaches,
is that the  dependence of the exchange terms from the angular momentum is
almost identical in the HFB$_{s+}$ and in the  HFB$_{e}$ approaches. 
For $I = 0 \hbar$  the binding energies differ only by about 500 keV.
The exchange terms are rather different in both approaches, however,  
when the  $\gamma$-ray  are calculated, the differences cancel  and one
gets similar values in both approaches.   
 
\begin{center}
\begin{table}
\begin{tabular}{cccccccc}
\hline
$I$  & App. &  $E_s$     &  $V^F_C$  &  $V^P_C$  &  $V^P_{SO}$ &   
$V^P_{TK}$  &   $E_{tot}$ \\
\hline
0           & $s$ &  -1766.259 &  -36.450  &  0.698    &   0.111     &   0.173       &   -1801.725 \\ 
            & $e$ &  -1765.862 &  -36.703  &  0.249    &  -0.007     &   0.091       &   -1802.232 \\
\hline
\hline
          &    & $\Delta E_s$  & $\Delta V^F_C$  &  $\Delta V^P_C$  & $\Delta V^P_{SO}$ & 
$\Delta V^P_{TK}$  &  $\Delta E_{tot}$ \\
\hline 
2  & $s$ &  42.5      &   -1.8  &       -4.5   &  -1.6	&   -1.7     &  32.9 \\
   & $e$ &  42.0      &   -1.5  &       -3.4   &  -2.7	&   -1.1     &	33.3 \\
\hline
4  & $s$ &  98.0      &   -4.7  &      -11.5   &  -4.2	&   -4.5     &  73.3 \\
   & $e$ &  99.1      &   -4.2  &       -9.4   &  -5.8	&   -2.6     &	77.1 \\
\hline
6  & $s$ & 152.0      &   -7.0  &      -17.4   &  -6.0	&   -6.5     & 114.8 \\
   & $e$ & 153.7      &   -6.8  &      -14.9   &  -8.6	&   -3.9     & 119.5 \\
\hline
8  & $s$ & 203.3      &   -9.5  &      -23.5   &  -8.0	&   -8.7     & 153.7 \\
   & $e$ & 207.8      &  -10.3  &      -22.2   & -10.4  &   -5.2     & 159.7 \\
\hline
10 & $s$ & 250.8      &  -11.6  &      -28.9   &  -9.2	&  -10.0     & 190.9 \\
   & $e$ & 257.4      &  -14.0  &      -29.1   & -11.4  &   -6.2     & 196.8 \\
\hline
12 & $s$ & 294.5      &  -14.6  &      -35.9   & -10.5	&  -11.9     & 221.4 \\
   & $e$ & 306.1      &  -19.5  &      -38.8   & -10.8  &   -7.0     & 229.9 \\
\hline
14 & $s$ & 333.7      &  -16.4  &      -40.9   & -10.9	&  -12.6     & 252.8 \\
   & $e$ & 359.0      &  -29.7  &      -55.7   &  -8.3  &   -7.8     & 257.6 \\
\hline
\end{tabular}
\caption{ Upper part: Ground state contributions to the binding energy of the
 nucleus  \nuc{240}{Pu} in the  HFB$_{s+}$ and in the  HFB$_{e}$ approaches.
 The energies are given  in MeV and the angular momenta in $\hbar$. Lower
 part: Contributions to the gamma ray energies along the Yrast band in keV
  in the  HFB$_{s+}$ and in the  HFB$_{e}$ approaches. }
\label{tab:240Pu}
\end{table}
\end{center}

\section{Superdeformed nuclei in the A $\approx 190$ region}
\label{sec:SHg}

\begin{figure}[h]
\begin{center}
\parbox[c]{14.cm}{\includegraphics[width=14.cm]{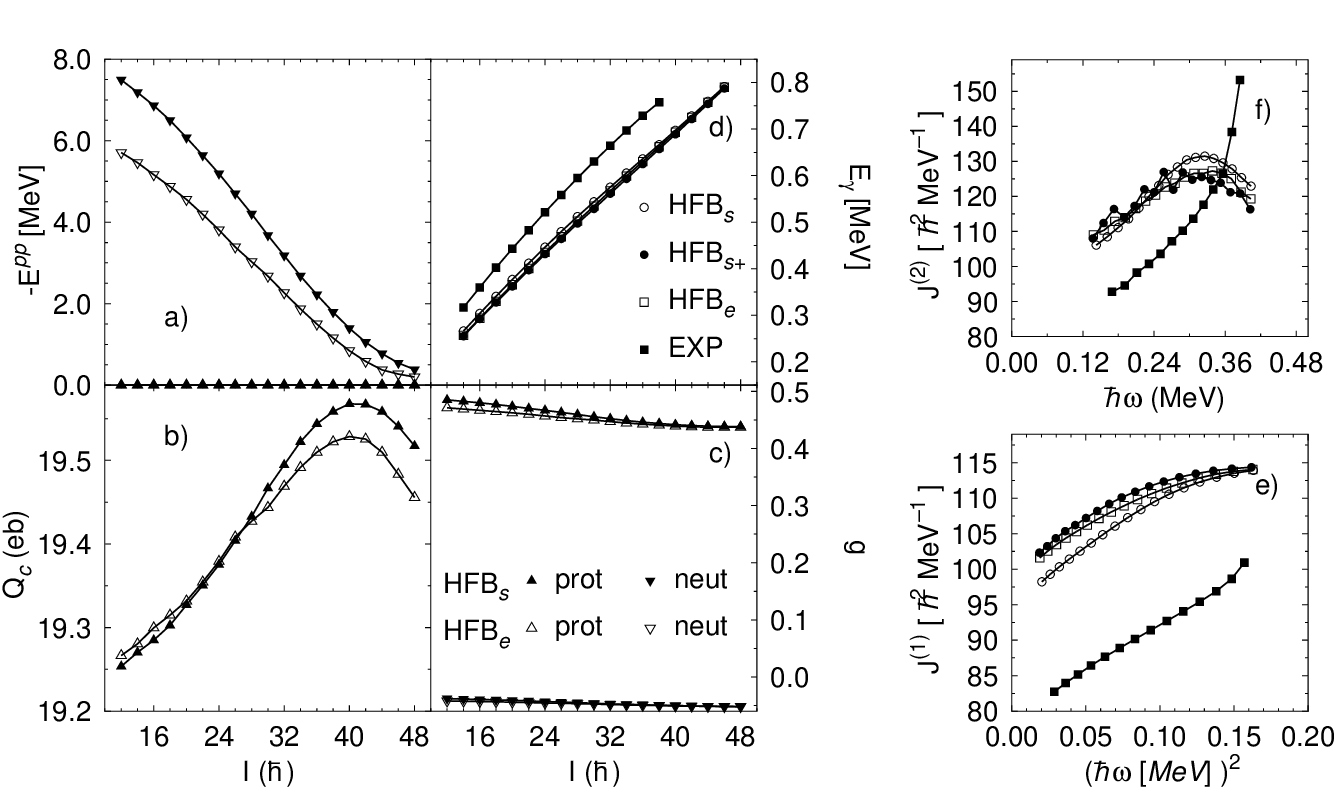}}
\end{center}
\caption{ Same as fig.~\ref{fig:Pu} but for the nucleus \nuc{190}{Hg}.
 The experimental data are from ref.\cite{190Hg}.}
\label{fig:Hg}
\end{figure}

This region of superdeformation has already been studied using the HFB$_s$ 
method with the Gogny force D1S~\cite{Gir94,VER.00}. In our calculations
we use the basis 
q$=1.5$, p$=1.0$ and N$_0=12.5$ and the D1S interaction. Taking into account
the results obtained in the other regions, we do not expect a big influence 
of the Coulomb exchange terms  because for the superdeformed states the
proton pairing energies either vanish  or are very small even with the 
HFB$_s$ method.  A large change of the pairing energies 
due to the Coulomb exchange and the other terms cannot exist, therefore,
in this states.  As an example of superdeformation we have chosen the nucleus 
\nuc{190}{Hg}. From this example we will learn  how the
other pairing contributions (spin--orbit and two--body correction) 
affect  the neutron pairing energies and the other nuclear properties through
the  selfconsistency. As before, we shall first look for the wave
function content of the HFB$_s$ and the HFB$_e$ approaches. In Fig.~\ref{fig:Hg}, 
panel (a),  we can see 
the behavior of the pairing  energies in both methods. As mentioned, the proton
 pairing energies along the band  vanish in both calculations. The neutron 
pairing energies  are close to each other in both calculations, though 
somewhat smaller values are obtained for the  HFB$_e$ at low angular momentum.
The charge quadrupole moment is depicted in panel (b). Only small differences
between both approaches are found at large angular momentum. The gyromagnetic
factors, panel (c), are again very close to each other. 
We shall now turn to a comparison between the  HFB$_s$,  HFB$_{s+}$ and HFB$_e$ predictions
for the $\gamma$-ray energies, panel (d), first moments of inertia, panel (e) and
second moments of inertia, panel (f).  For these observables we do not find remarkable differences
between the three approaches.
From this section we may conclude that the Coulomb exchange and pairing
contribution barely modify the properties of the nucleus when the
approximate calculation HFB$_s$ has vanishing proton pairing energies. 
Then, the approach HFB$_s$ is a good approximation to the completely
self--consistent calculation, HFB$_e$, even for the wave function. 

\section{ Energy Surfaces}
\label{sec:enersurf}
     
\begin{figure}[h]
\begin{center}
\includegraphics[width=8.cm,angle=270]{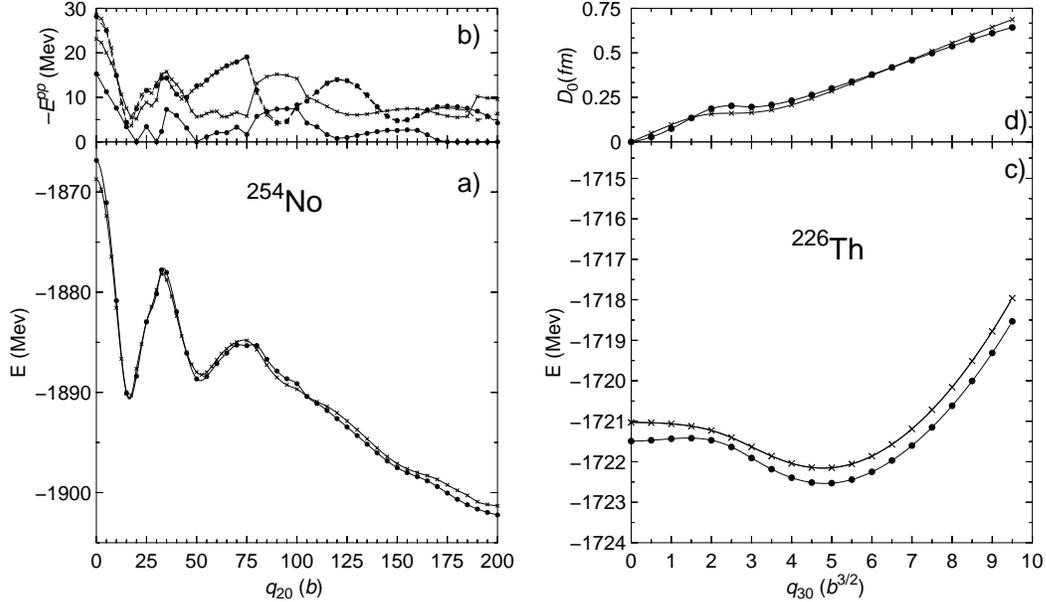}
\end{center}
\caption{a) Results in the HFB$_{s,Sla}$ approach, crosses, and
HFB$_{e,nso}$, filled circles.
a)  Binding energy of the nucleus \nuc{254}{No} versus the quadrupole moment.
 b) Pairing energies along the fission barrier, protons (neutrons), 
dashed (solid) line.
c)  Binding energy of the nucleus \nuc{226}{Th} versus the octupole moment.
d) Dipole moment versus the octupole moment.}
\label{fig:enesurf}
\end{figure}

  In this section we shall finally investigate the effect of the exchange
terms as a function of a collective variable at zero angular momentum. We have
used an axially symmetric HFB code that allows reflexion asymmetric shapes.
We have performed two different calculation. In the first one we have
neglected all exchange terms, i.e., the HFB$_s$ approach, but we have included, 
the Fock Coulomb term in the Slater approximation, we shall call this approach
HFB$_{s,Sla}$. The second approach includes all exchange terms with the 
exception of the contribution of the spin-orbit term to the pairing 
field,\footnote{We do not include this term in the calculations just because 
it is not relevant for the problem we are considering, besides the fact
that it is very cumbersome to include in the axial HFB code.} we shall call 
this approach
HFB$_{e,nso}$. Notice that since we do not add at the end of the calculation
HFB$_{s,Sla}$ the pairing terms of the Coulomb and the two-body kinetic 
energy (both contributions are repulsive) it is possible to find HFB$_{s,Sla}$ 
solutions with an energy deeper than the  HFB$_{e,nso}$ approach. That means
we are more interested in the general behavior of the energy surface than in 
the absolute values. 
The most common case of energy dependence as a function of a collective
variable appears in the calculation of fission barriers. 
In this calculations we use the basis q$=1.5$, p$=1.0$ and N$_0=15.1$ and the 
D1S parameterization. 
In fig.~\ref{fig:enesurf}, on panel a), we display the binding energy of 
\nuc{254}{No} as a function of the constrained quadrupole moment in both
approaches.
We find a similar behavior in both calculations, though small differences are
noticeable. For example, the height of the first barrier is slightly 
larger in the  
HFB$_{e,nso}$ approach than in the HFB$_{s,Sla}$. The second one, however, 
has more or less the same height in both approaches. In panel b) we display
the pairing energies along the fission path.  The neutron pairing energy
is practically the same in both calculations. The proton pairing energies 
in the
HFB$_{e,nso}$ calculations show the Coulomb antipairing effect already
discussed in other nuclei. 
They are in absolute value 5 to 10 MeV smaller than in the calculation without
Coulomb pairing term.

We shall now investigate the effect of the exchange terms in cases where
the reflexion symmetry is broken i.e. the wave functions do not have a good
parity.  In these calculations we use the same theoretical approaches as in
the  \nuc{254}{No} ones, we use the basis q$=1.3$, p$=1.0$ and N$_0=13.1$ 
and the D1S interaction. 
In panel c) of the same figure, we present the binding energy of the nucleus 
\nuc{226}{Th} versus the constrained octupole moment. As with the quadrupole
operator we do not find any remarkable difference between both calculations.
The depth and shape of the octupole deformed minimum are almost identical
in both approaches. In panel d), finally, we present the dipole moment versus 
the constrained octupole momentum.  Again, small differences between both
calculations are found but without further relevance.

\section{Discussion}

The main outcome of our investigation is that the exchange terms usually
neglected affect mainly the nuclear pairing properties. More specifically,
they affect mostly the {\it proton} nuclear pairing properties because 
of all missing terms the largest contribution to the pairing field arises 
from the Coulomb potential.
Observables little or not dependent on the pairing correlations, are, therefore,
not  affected by these terms. On the contrary, observables like the moments
of inertia or two-quasiparticle states energies, strongly dependent on pairing
 correlations, are very sensitive to these terms.

We have found that the results of calculations performed with the plain HFB$_s$
approximation differ significantly, for some observables, from those of the 
exact solution HFB$_e$, 
in nuclei where the proton pairing energies are relatively large. On the 
contrary, the HFB$_{s+}$ results agree, in general, very well with those 
of the exact solution. In cases of very small proton correlation energies all 
three approaches, HFB$_s$, HFB$_{s+}$ and  HFB$_e$ practically coincide. 

 For those situations where proton correlations are relevant,
the predictions of the HFB$_s$ approximation  agree 
better with the experimental data than the exact ones or than the ones from
the HFB$_{s+}$ approximation. Paradoxically, the Gogny force was fitted 
\cite{Dec80} using the {\it exact} HFB approach to calculate properties of
spherical nuclei. 
The pairing properties of the Gogny force were adjusted by fitting
the odd-even mass difference of the $Sn$ isotones. The $Sn$ isotones, however,
have a major shell closure in protons, Z$=50$, and in this case, 
practically, it does not matter whether one performs approximate or exact 
calculations, i.e., effectively the Coulomb exchange terms were not considered
in this fit. Furthermore, the fact that including the exchange terms in the
calculations worsens the good agreement of the theoretical results with the 
experimental data, reinforces the conjecture that the actual fits of the Gogny 
force should be used,  in calculations with open proton shells, without the
mentioned terms.

 Of course the Gogny force is isospin invariant but the Coulomb force is 
not. The Hartree-Fock potential that one obtains after solving the HFB equations
neither is isospin invariant. It is not clear, therefore, that a different
parameterization (concerning the pairing part of the interaction) of the force
would not be obtained if nuclei with open proton  shells had been considered in the 
fit.   The whole issue of adjusting the pairing properties of any force
is itself rather fuzzy. Most of the fits have been done to reproduce the
experimentally observed odd-even mass difference. This energy difference 
can, however, not be attributed only to  pairing correlations 
\cite{DMN.00}. From the theoretical point of view
the calculation of this energy difference presents also problems because
binding energies of odd nuclei are not easy to calculate. Blocking effects
and angular momentum projection should be performed in order to evaluate
properly this energy, besides additional couplings. However, in order to save
CPU time during the fit, approximations are used, in particular in order not 
to break the time reversal symmetry the equal filling approximation is used.

Taking into account the complication of the neglected exchange terms, 
see Appendix~B, 
 the fact that the HFB$_s$ approximation, commonly used in most HFB calculations, 
provides such good agreement with the experimental data is 
very fortunate.

\section{Conclusions}

 For the first time we have performed an exact self--consistent HFB calculation
in a triaxial basis with the Gogny force.
 The exact Coulomb exchange term and the contributions of the spin--orbit, 
two--body correction of the kinetic energy and Coulomb terms to the pairing 
field have been included in the calculations. An exhaustive analysis of each 
term has been performed for spherical nuclei in different mass regions. We find 
the Coulomb contribution to the pairing field to be the most relevant one.
The other terms, though not negligible, give  smaller contributions.  

For deformed nuclei we have studied the commonly used HFB$_s$ approximation,
the perturbative HFB$_{s+}$, and the exact solution. We have been concerned 
with different mass regions at zero and high spins. As in the case of spherical
 nuclei, 
we observe a reduction of the proton pairing energies in the exact calculations
as compared to the HFB$_s$ ones. Ground state energies are very similar
in the perturbative and in the exact calculations. Energies of excited states
are somewhat different in the HFB$_s$ approximation than in the exact one.
We have also investigated superdeformed states. Since the proton pairing
correlations of these nuclei are small we do not find any significant difference 
between the approximate and the exact solution.
Lastly, we have analyzed the effect of the most relevant exchange terms in the
calculation of energy surfaces. The energy as a function of the quadrupole
(octupole) mass operator is practically identical in the approximate and in
the exact calculations.  
 
 We also have found, that the Slater approximation for the Fock Coulomb term is a
good one for all HFB calculations,  in agreement with ref.\cite{Tit74}
for the HF case.

 For all nuclei and situations analyzed the following general conclusions apply:

\begin{enumerate}

\item In cases where the proton pairing correlations do not play an important
role, anyone of the approximations considered practically coincide with the 
exact calculation. 

\item When proton pairing correlations are relevant, then:
\begin{itemize}
\item   For energy-related observables the perturbative approach
HFB$_{s+}$ is a good approximation to the exact one.
\item For those observables whose evaluation requires the wave functions,
one should perform the exact calculation.
\item If the HFB wave functions are thought as a basis for more elaborated
theories, like RPA or GCM, then the exact calculation is required.
\end{itemize}  
\item The way in which the fits of the Gogny force were performed favors
neglecting the mentioned exchange terms
 in HFB calculations using these fitted parameter sets. 
\end{enumerate}

It remains to investigate the real importance of this terms by performing 
new fits of the Gogny force taking into account explicitly all terms
discussed here.

We thank J.F. Berger, M. Girod and S. Peru for discussions.
This work was supported in part by DGICyT, Spain under Project PB97-0023.
One of us (M.A) acknowledges a grant from the Spanish Ministery of Education
under Project PB94-0164.

\newpage
\appendix
{Appendix A: The interaction}

 As an effective  interaction we use the Gogny  force \cite{Dec80}
\begin{eqnarray}
v_{12} & = & \sum_{i=1}^2 e^{-{(\vec{r}_1 - \vec{r}_2)}^2/\mu_i^2} (W_i +
B_iP_{\sigma} -H_i P_{\tau} -M_i P_{\sigma} P_{\tau} ) + \nonumber \\
& + & W_{LS} (\vec{\sigma}_1 + \vec{\sigma}_2) \vec{k} \times \delta(\vec{r}_1 -
\vec{r}_2) \vec{k} + \nonumber \\
& + & t_3(1+x_0 P_{\sigma}) \delta (\vec{r}_1 -\vec{r}_2) \rho^{1/3} \left (
\frac{1}{2} (\vec{r}_1 + \vec{r}_2 )\right ),
\label{eq:vgog}
\end{eqnarray}
and the Coulomb force
\begin{equation}
v_{12}^C = (1+2\tau_{1z})(1+2\tau_{2z}) \frac{e^2} {|\vec{r}_1-\vec{r}_2 | }.
\end{equation}
Additional contributions taken into account in the calculations arise from
the one--body and two--body  center of mass corrections 
\begin{equation}
\hat{T}  =  \sum_i \frac{{\vec{p}_i}^{\,2}}{2m} \left ( 1 -\frac{1}{A} \right ) -
\frac{1}{Am} \sum_{i>j} \, \vec{p}_i \cdot \vec{p}_j.
\end{equation}
In the calculations we use the parametrizations D1S \cite{Ber91}
 and D1~\cite{Dec80,Gog75}.

\appendix
{Appendix B: Calculation of the Coulomb Hartree-Fock field and
 pairing tensor.}

To compute the Hartree-Fock field and pairing tensor for the Coulomb
interaction we have followed the idea used in \cite{Gir83}. It amounts to use the
identity  
\[ \frac{1}{|\vec{r}_{1}-\vec{r}_{2}|}=\frac{2}{\sqrt{\pi }} \int
_{0}^{\infty }\frac{d\mu }{\mu ^{2}} e^{-\left( \vec{r}_{1}-\vec{r}_{2}\right)
^{2}/\mu ^{2}},\]
to compute the Coulomb matrix elements in terms of those of
Gaussians with range \( \mu  \). The main advantage of this approach is that
the matrix elements of the Gaussians are factorizable in terms of quantities
defined for each Cartesian direction \( x \), \( y \) and \( z \).
Unfortunately, the price to pay is the \( \mu  \) integration. Introducing the
HF field \( \Gamma ^{G}_{k_{1}k_{2}}(\mu ) \) and the pairing tensor \( \Delta
^{G}_{k_{1}k_{2}}(\mu ) \), see eq.(\ref{fields}), of a Gaussian of range
\( \mu  \) we can express the corresponding quantities for the Coulomb 
potential as
 \[ \Gamma
^{C}_{k_{1}k_{2}}=\frac{2e^{2}}{\sqrt{\pi }}\int _{0}^{\infty } \frac{d\mu
}{\mu ^{2}}\Gamma ^{G}_{k_{1}k_{2}}(\mu )\] 
and 
\[ \Delta
^{C}_{k_{1}k_{2}}=\frac{2e^{2}}{\sqrt{\pi }}\int _{0}^{\infty } \frac{d\mu
}{\mu ^{2}}\Delta ^{G}_{k_{1}k_{2}}(\mu )\]
 To transform the integration interval to a finite one, one makes the change of 
variables
\( u^{2}=\left( 1+\alpha \mu ^{2}\right) ^{-1} \), with \( \alpha  \) a
parameter to be determined below.  We finally obtain 
\[ \Gamma
^{C}_{k_{1}k_{2}}=2e^{2}\sqrt{\frac{\alpha }{\pi }}\int
_{0}^{1}\frac{du}{\left( 1-u^{2}\right) ^{3/2}}\Gamma ^{G}_{k_{1}k_{2}}\left(
\frac{\sqrt{\frac{1}{\alpha }\left( 1-u^{2}\right) }}{u}\right) \]
 and 
\[
\Delta ^{C}_{k_{1}k_{2}}=2e^{2}\sqrt{\frac{\alpha }{\pi }}\int
_{0}^{1}\frac{du}{\left( 1-u^{2}\right) ^{3/2}}\Delta ^{G}_{k_{1}k_{2}}\left(
\frac{\sqrt{\frac{1}{\alpha }\left( 1-u^{2}\right) }}{u}\right) .\] 
The integrations are carried out numerically with the Gauss-Legendre method.
 The optimal choice of \( \alpha  \), turns out to be  equal to the
maximum of \( \left( \alpha _{x},\alpha _{y},\alpha _{z}\right)  \) where \(
\alpha _{i}=\frac{1}{2b_{i}^{2}} \) and \( b_{i} \)
are the oscillator lengths (see ref.~\cite{Gir83}).

In the numerical computation of the Coulomb exchange and pairing terms we have
used 20 points for the Gauss-Legendre integration. This number of points has
proven to be enough for a precise determination of the exchange field and
pairing tensor. The number of integration points used makes the evaluation of
the Coulomb terms rather costly in terms of CPU time: it usually takes a factor
5.5 longer to compute those factors than to calculate the Fock and
pairing parts of the Brink-Boecker interaction (in this case, we have two
gaussians and we have to compute them for both protons and neutrons). The
calculation of the exchange and pairing parts of the BB interaction is already
the most costly part of each iteration as it usually takes between a factor 10
and 15 longer than the evaluation of the direct terms. 
In a standard workstation
and for a configuration space with parameters \( p=1 \), \( q=1.3 \) and \(
N_{0}=11.1 \) the evaluation of the direct field takes 1.9 seconds,
 the combined Fock and pairing terms of the BB part takes 19 seconds 
 and the evaluation of the Fock and pairing parts of the Coulomb interaction 
takes 103 seconds. One has to keep
in mind that the previous numbers are for each iteration of the minimization
process. If we increase the size of the basis by using \( q=1.6 \) and \(
N_{0}=14.0 \) the previous numbers climb up to 7, 92 and 540 seconds
respectively. We clearly see the tremendous slow down introduced in the
calculations stemming from the consideration of the Fock and pairing terms
 of the
Coulomb interaction. This tremendous slow down can be even worse for the
practitioners of the Skyrme interaction as their CPU time consumption per
iteration has to be of the order of the one needed to calculate the direct
 field of the Gogny force.

Concerning the computation of the pairing field associated to the spin-orbit
interaction we have followed the ideas already presented in \cite{Gir83}. An 
explicit expression for all these terms can be found in \cite{Ma.00}. The
 numerical computation of this field takes almost the same time as the 
 computation of the HF field and therefore, its inclusion is not significant 
 in terms of CPU time.

\end{document}